\pdfoutput=1

\documentclass[11pt,twoside,a4paper,cmspaper,final,collab]{cms-tdr}
\def\svnVersion{492824P}\def\cmsCernNoTag{CERN-EP-2018-258}\def\cmsCernDate{\today}\def\cmsMessage{Published in the Journal of High Energy Physics as \href{http://dx.doi.org/10.1007/JHEP03(2019)082}{\doi{10.1007/JHEP03(2019)082.}}}

\begin{document}\cmsNoteHeader{B2G-17-014}

\hyphenation{had-ron-i-za-tion}
\hyphenation{cal-or-i-me-ter}
\hyphenation{de-vices}
\RCS$HeadURL: svn+ssh://svn.cern.ch/reps/tdr2/papers/B2G-17-014/trunk/B2G-17-014.tex $
\RCS$Id: B2G-17-014.tex 492798 2019-03-26 16:33:12Z sisagir $

\newlength\cmsTabSkip\setlength{\cmsTabSkip}{1ex}

\newcommand{\xft}{\ensuremath{X_{5/3}}\xspace}
\newcommand{\HTl}{\ensuremath{\HT^{\text{lep}}}}
\newcommand{\drlj}{\ensuremath{\Delta R(\ell\mathrm{,j_{2})}}\xspace}
\newcommand{\minmlb}{\ensuremath{\min[M(\ell,\cPqb)]}\xspace}
\newcommand{\minmlj}{\ensuremath{\min[M(\ell,\mathrm{j})]}\xspace}
\newcommand{\nconst}{\ensuremath{N_{\text{const}}}}
\newcommand*{\MADSPIN}{\textsc{MadSpin}\xspace}
\providecommand{\cmsTable}[1]{\resizebox{\textwidth}{!}{#1}}
\providecommand{\NA}{\ensuremath{\text{---}}}
\providecommand{\CL}{CL\xspace}

\cmsNoteHeader{B2G-17-014}
\title{Search for top quark partners with charge 5/3 in the same-sign dilepton and single-lepton final states in proton-proton collisions at $\sqrt{s} = 13\TeV$}

\date{\today}

\abstract{
A search for the pair production of heavy fermionic partners of the top quark with charge 5/3 (\xft) is performed in proton-proton collisions at a center-of-mass energy of 13\TeV with the CMS detector at the CERN LHC.
The data sample analyzed corresponds to an integrated luminosity of 35.9\fbinv.
The \xft quark is assumed always to decay into a top quark and a {\PW} boson.
Both the right-handed and left-handed \xft couplings to the {\PW} boson are considered.
Final states with either a pair of same-sign leptons or a single lepton are studied.
No significant excess of events is observed above the expected standard model background.
Lower limits at 95\% confidence level on the \xft quark mass are set at 1.33 and 1.30\TeV respectively for the case of right-handed and left-handed couplings to {\PW} bosons in a combination of the same-sign dilepton and single-lepton final states.
}

\hypersetup{
pdfauthor={CMS Collaboration},
pdftitle={Search for top quark partners with charge 5/3 in the same-sign dilepton and single-lepton final states in proton-proton collisions at sqrt(s) = 13 TeV},
pdfsubject={CMS},
pdfkeywords={CMS, physics, top quark partners}}

\maketitle

\section{Introduction}

The prediction of new heavy quarks is a common feature of many theories of physics beyond the standard model (SM).
In composite Higgs models~\cite{Contino:2008hi,TopPartnerHunterGuide,Pomarol}, heavy partners of the SM top quark solve the hierarchy problem caused by quadratic divergences in the quantum-loop corrections to the Higgs boson mass by providing contributions that offset those due to the SM top quark.
Often in such models, new color-triplet partners are predicted, with one of them having an exotic electric charge of 5/3 times the charge of the positron, referred to as \xft.
In partially composite scenarios~\cite{Kaplan}, these exotically charged fermions need not contribute to the gluon-gluon fusion production mode of the Higgs boson~\cite{Azatov:2011qy} and hence such measurements set no constraints on the mass of the \xft particle.
This paper describes a search for such a fermionic top quark partner, using proton-proton ($\Pp\Pp$) collision data collected during 2016 at a center-of-mass energy of 13\TeV with the CMS experiment at the CERN LHC, corresponding to an integrated luminosity of 35.9\fbinv.

\begin{figure}[hbtp]
\centering
\includegraphics[width=0.49\textwidth]{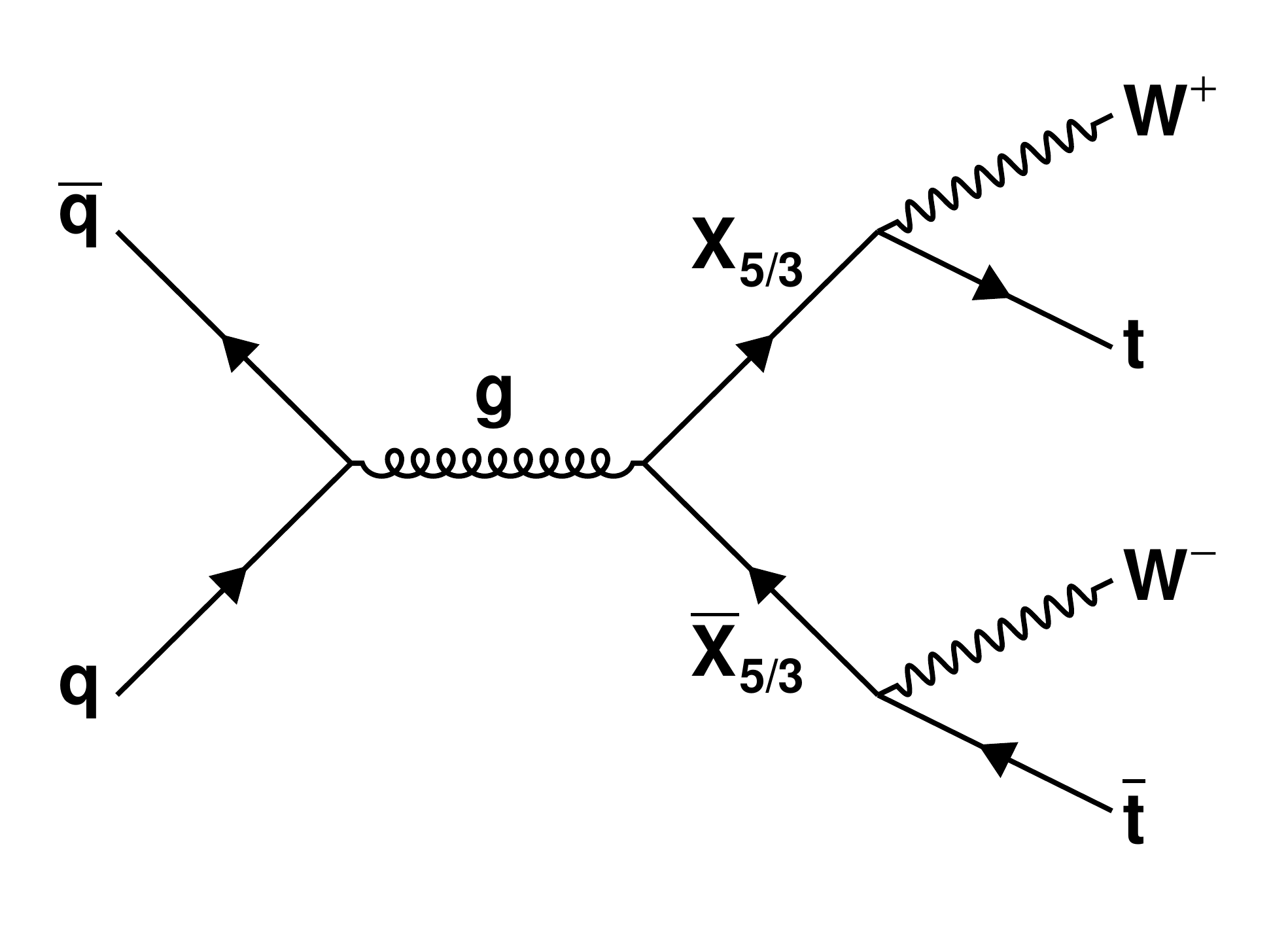}
\includegraphics[width=0.49\textwidth]{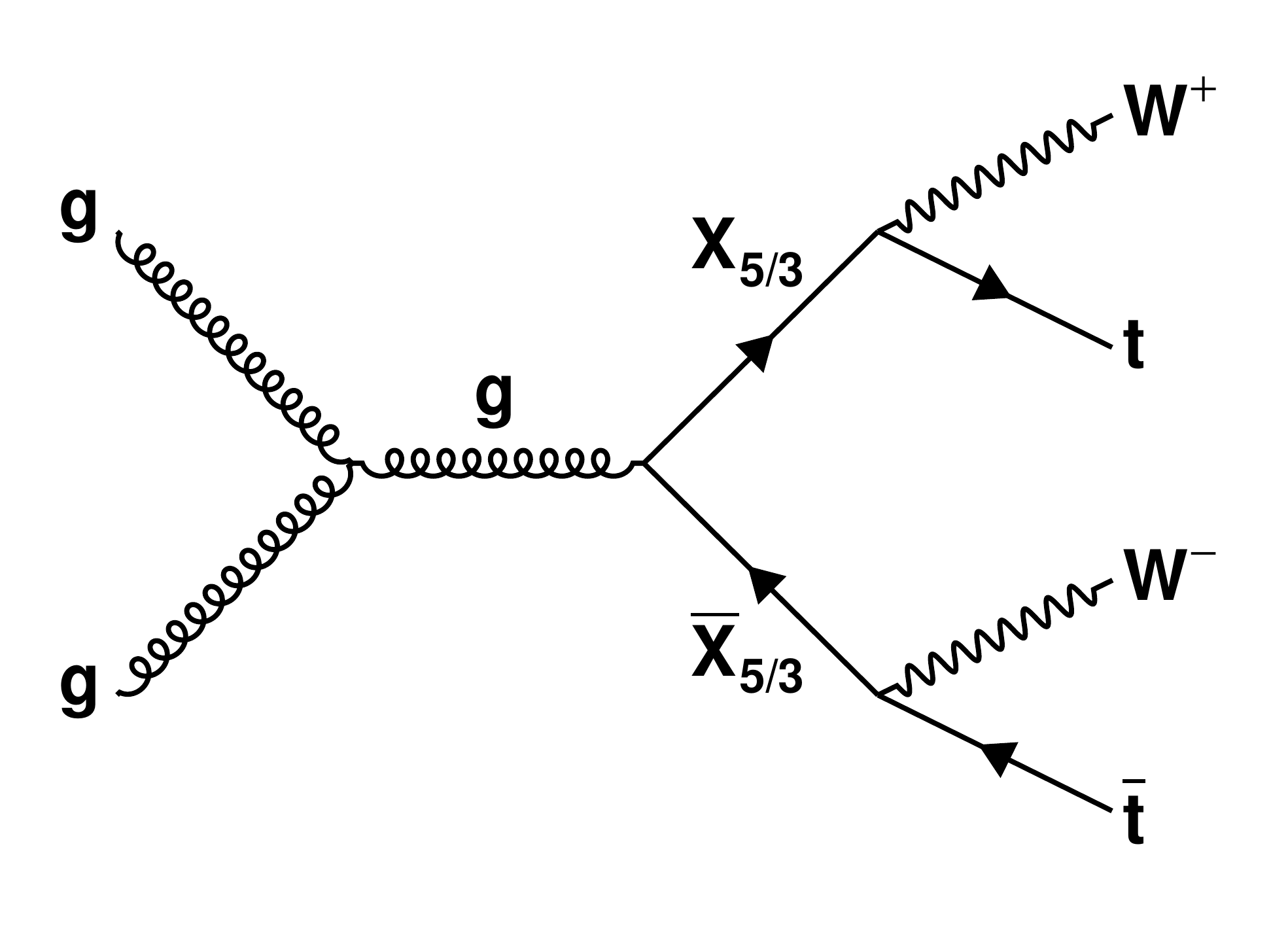}
\caption{Leading order Feynman diagrams showing pair production and decays of \xft particles via QCD processes.}
\label{fig:feynman}
\end{figure}

The dominant mechanism for \xft production, shown in Fig.~\ref{fig:feynman}, is via quantum chromodynamics (QCD) processes, which yield particle-antiparticle pairs, since the \xft carries color charge.
The \xft particle can also be singly produced via electroweak processes, but that production mode is model dependent and is not considered here.
Since the pair production involves exclusively the SM QCD coupling, the tree-level cross section is independent of the \xft properties, other than its mass.
The \xft particle is assumed to decay into a top quark and a {\PW} boson with a branching fraction of 100\%, since this is the dominant decay mode in many models~\cite{Cacciapaglia}.
The decay can occur through either right-handed (RH) or left-handed (LH) couplings to {\PW} bosons, and this search presents results for either fully RH or fully LH decays.
Thus we have not restricted the interpretation of the results to the case of vector-like quarks, whose left-handed and right-handed chirality states have the same transformation properties under the weak isospin $\mathrm{ SU(2) }$ gauge group, although limits obtained with this assumption would be very similar to those set for pure-LH or pure-RH couplings.

This search focuses on two different final states consisting of either exactly one lepton or multiple leptons with the requirement that there be a pair of same-sign leptons.
In both cases, additional hadronic activity in the event is required.
Throughout the paper, the word lepton refers to an electron or a muon.
Although leptonic tau decays are not specifically targeted in this analysis, their contribution to the signal efficiency is taken into account.
The same-sign dilepton final state relies on its relatively clean signature and the large amount of jet activity from the other \xft particle in the event to discriminate against background processes.
The single-lepton channel exploits the shape of the distribution of the visible mass of the top quark reconstructed in the detector to discriminate against background events.

Previously, CMS conducted a search for the \xft particle using data collected at a center-of-mass energy of 8\TeV, corresponding to an integrated luminosity of 19.5\fbinv, in the same-sign dilepton channel only, setting a lower limit on the \xft mass of 800\GeV at 95\% confidence level (\CL)~\cite{8TeVPaper}.
CMS has recently carried out another search~\cite{B2G-15-006} for \xft in a combination of the same-sign dilepton and single-lepton final states using data collected in 2015 at $\sqrt{s} = 13\TeV$, corresponding to an integrated luminosity of 2.3\fbinv, setting a lower limit on the \xft mass of 1.02 (0.99)\TeV for an RH (LH) coupling.
Searches have also been performed by the ATLAS experiment at center-of-mass energies of 8 and 13\TeV~\cite{Aad:2015gdg,Aad:2015mba,Aaboud:2017zfn,Aaboud:2018uek,Aaboud:2018xpj,Aaboud:2018pii}. The results based on $\sqrt{s} = 13\TeV$ with 36.1\fbinv of data set a lower limit of 1.37\TeV on the mass of the \xft particle.

The present search follows closely the strategy of Ref.~\cite{B2G-15-006} and benefits from an order of magnitude increase in the integrated luminosity.
This paper is organized as follows: Section~\ref{sec:CMS} briefly describes the CMS detector;
Section~\ref{sec:Simulation} discusses the simulated signal and background samples;
in Section~\ref{sec:Triggering}, trigger details are given;
Section~\ref{sec:Reconstruction} contains a description of the event reconstruction;
the analyses of the same-sign dilepton and single-lepton final states are detailed in Sections~\ref{sec:SSDL}--\ref{sec:LJets};
and the systematic uncertainties are discussed in Section~\ref{sec:Systematics}.
Finally, Sections~\ref{sec:Results}--\ref{sec:Summary} give the results and a summary.

\section{The CMS detector}\label{sec:CMS}

The central feature of the CMS apparatus is a superconducting solenoid of 6\unit{m} internal diameter, providing a magnetic field of 3.8\unit{T}. Within the solenoid volume are a silicon pixel and strip tracker, a lead tungstate crystal electromagnetic calorimeter (ECAL), and a brass and scintillator hadron calorimeter (HCAL), each composed of a barrel and two endcap sections. Forward calorimeters extend the pseudorapidity ($\eta$) coverage provided by the barrel and endcap detectors. Muons are detected in gas-ionization chambers embedded in the steel flux-return yoke outside the solenoid.
Events of interest are selected using a two-tiered trigger system~\cite{Khachatryan:2016bia}. The first level, composed of custom hardware processors, uses information from the calorimeters and muon detectors to select events at a rate of around 100\unit{kHz} within a time interval of less than 4\mus. The second level, known as the high-level trigger, consists of a farm of processors running a version of the full event reconstruction software optimized for fast processing, and reduces the event rate to less than 1\unit{kHz} before data storage.

A more detailed description of the CMS detector, together with a definition of the coordinate system used and the relevant kinematic variables, can be found in Ref.~\cite{Chatrchyan:2008zzk}.

\section{Simulation}\label{sec:Simulation}

{\tolerance=400
The \xft signal processes are generated using a combination of
\MGvATNLO 2.2.2~\cite{Alwall:2014hca} and \MADSPIN~\cite{MadSpin}
for two coupling scenarios: allowing only RH or only LH \xft coupling to {\PW} bosons.
The \MGvATNLO event generator is used both to produce \xft events and to decay each \xft to a top quark and a {\PW} boson, while the decays of the top quarks and {\PW} bosons are simulated with \MADSPIN.
The signal events are simulated at leading order (LO) for \xft masses from 800 to 1500\GeV, in 100\GeV steps, separately for each coupling scenario.
The signal samples are then normalized to the next-to-next-to-leading order cross sections using the \textsc{Top++}2.0 generator~\cite{TPRIMEXSEC,MITOV1,MITOV3,MITOV2,BARNREUTHER,NNLL}, with resummation of soft gluon corrections at the next-to-next-to-leading logarithmic accuracy.
\par}

A variety of event generators are used for the Monte Carlo (MC) simulation of the background processes.
The \POWHEG 2.0~\cite{Nason:2004rx,Frixione:2007vw,Alioli:2010xd,Frixione:2007nw} event generator is used to simulate {\ttbar}, single top quark events in the \textit{t}-channel and {\cPqt}{\PW} channel, {\ttbar}{\PH}, {\PW}{\PZ}, and {\PZ}{\PZ} events to next-to-leading order (NLO) precision.
The \MGvATNLO event generator is used to simulate {\PZ}+jets, {\PW}+jets, single top quark process in the \textit{s}-channel, {\ttbar}{\PZ}, {\ttbar}{\PW}, and {\ttbar}{\ttbar} processes, events with a combination of three {\PW} or {\PZ} bosons, and QCD multijet events.
The {\PZ}+jets, {\PW}+jets, \PW\PW, and QCD multijet processes are generated at LO using the MLM matching scheme~\cite{Alwall:2007fs}. The FxFx matching scheme~\cite{Frederix:2012ps} is used for {\ttbar}{\PZ}, {\ttbar}{\PW}, {\ttbar}{\ttbar}, triboson, and single top quark process in the \textit{s}-channel, which are generated at NLO.

Additional {\Pp\Pp} interactions in the same or neighboring bunch crossings (pileup) are modeled by superimposing simulated minimum-bias interactions onto the simulated events for all processes.
Simulated events are reweighted so that the number of pileup interactions matches the distribution observed in data.

{\tolerance=800
Parton showering, hadronization, and the underlying event are simulated with \PYTHIA 8.212~\cite{PYTHIA82}, using NNPDF 3.0~\cite{NNPDF30} parton distribution functions (PDFs) and the CUETP8M1 tune~\cite{CUETP8M1,Skands:2014pea} for all MC processes, except for the \ttbar sample, which is produced with the CUETP8M2T4 tune~\cite{CMS-PAS-TOP-16-021}.
Finally, for all MC samples, generated events are processed through the full \GEANTfour-based simulation of the CMS detector~\cite{GEANT} and then reconstructed using the same procedure as the data.
\par}

The transverse momentum (\pt) spectrum of the top quarks in \ttbar events is known to be mismodeled in simulation~\cite{TopPt} and, therefore, corrections are applied to simulated \ttbar events as a function of the top quark \pt.

Many of the SM background processes in this search are similar and are therefore grouped together in the discussion that follows.
The same-sign dilepton final state groups SM processes according to their similarity to the signal topology and classifies them as ``{\ttbar}+X'',
containing {\ttbar}{\PW}, {\ttbar}{\PZ}, {\ttbar}{\PH}, and {\ttbar}{\ttbar}, which are those processes most similar to the signal, and ``multiboson'', comprising all processes mentioned above where two or three electroweak bosons are directly produced.
For the single-lepton final state, the background processes are grouped into three categories.
The first category is referred to as ``TOP'', which is dominated by \ttbar events, but also includes any process having at least one top quark.
The second category is referred to as ``EWK'', which is dominated by {\PW}+jets events, but includes all processes that contain electroweak bosons and no top quark.
The third category is referred to as ``QCD'' and is the QCD multijet background.

\section{Trigger and event selection}\label{sec:Triggering}

For the same-sign dilepton final state, candidate events are required to have passed triggers based on two electrons, two muons, or electron-muon combinations.
For the first half of the data set, symmetric trigger \pt thresholds were used for the dielectron and electron-muon triggers, corresponding to a \pt requirement of 33 (30)\GeV for the former (latter).
During the data-taking period, the instantaneous luminosity of the LHC steadily increased. 
Therefore, for the second half of the data set, to keep the trigger rate at an acceptable level, these triggers were replaced with new ones that had asymmetric \pt requirements, with the higher \pt (leading) lepton requirement of 37\GeV and the lower \pt (subleading) lepton threshold of 27\GeV, for both the dielectron and electron-muon triggers.
Throughout the entire data taking period, the same dimuon trigger, which had \pt requirements of 30 (11)\GeV for the leading (subleading) muon, was used.

In the single-lepton final state, events are required to pass either single-electron or single-muon triggers.
For the single-electron triggers, either an electron isolated from nearby particles with $\pt>32\GeV$, or a very loosely isolated electron with $\pt>15\GeV$ together with $\HT>350\GeV$ is required, where \HT is the scalar \pt sum of all jets at the trigger level with $\pt>30\GeV$ and $\abs{\eta}<3.0$.
The single-muon triggers require either a muon with $\pt>50\GeV$ with no isolation requirement or a very loosely isolated muon with $\pt>15\GeV$ together with $\HT>350\GeV$.

\section{Object reconstruction}\label{sec:Reconstruction}

This search makes use of electrons, muons, jets, and missing transverse momentum.
The reconstruction of these objects is based on a particle-flow (PF) algorithm~\cite{pf}, which reconstructs and identifies particles using an optimized combination of subdetector information.

The candidate events are required to have at least one reconstructed vertex passing basic quality criteria.
In the case that there are multiple reconstructed vertices, the one with the largest value of summed physics-object $\pt^2$ is taken to be the primary {\Pp\Pp} interaction vertex. Here, the physics objects are the jets, clustered using the jet finding algorithm~\cite{Cacciari:2008gp,FastJet3} with the tracks assigned to the vertex as inputs, and the associated missing transverse momentum, taken as the negative vector \pt sum of those jets.

Electron candidates are reconstructed from a collection of electromagnetic clusters that are matched to reconstructed tracks in the tracker~\cite{8TeV-EGamma}.
As in Ref.~\cite{B2G-15-006}, the identification criteria for electrons are based on a multivariate analysis (MVA), which makes use of shower shape variables,
track quality requirements, variables measuring compatibility between the track and matched
electromagnetic clusters, distance from the track to the primary vertex, and the probability that the electron candidate arises from a photon conversion.

In the same-sign dilepton final state, a consistency requirement is placed on the three measurements of the electron charge that result from three different methods.
Two of these charge assignment methods are based solely on tracker information, where the charge of the track is determined by the standard CMS track reconstruction~\cite{Khachatryan:1277738} or the Gaussian Sum Filter algorithm~\cite{gsf}. A third method is based on the difference in azimuthal angle ($\phi$) between the ECAL cluster center of gravity and pixel detector seeds used to reconstruct the electron track.
Because the third method has been found to be unreliable at high \pt, only the results from the first two charge determination methods are required to agree for electrons with \pt above 100\GeV .
Relaxing the requirement on this method recovers 5--10\% of signal efficiency, depending on the mass of the \xft.
For electrons with \pt below 100\GeV, all three charge measurements are required to agree.

Muons are reconstructed using a global track fit of hits in the muon chambers and hits in the silicon tracker.
The identification criteria are based on the number of hits used in the fit, the track quality, and the distance of the track to the primary vertex.
For the same-sign dilepton final state in the dimuon channel, the two muons should not be both within $\abs{\eta}>1.2$, unless they are in opposite sides of the detector in $\eta$ or are well separated in $\phi$ ($\Delta\phi > 1.25\unit{rad.}$).
This last requirement is imposed because of a misconfiguration of part of the trigger system, in the first part of the data-taking period, affecting nearby muons in the endcap detectors and has no effect on signal efficiency.

We select charged leptons that are isolated from other activity in the detector. The isolation variable ($I$) for both electrons and muons is defined as the scalar \pt sum of all PF candidates within a cone of varying size around the particle, divided by its \pt.
The radius used for the isolation cone ($\mathcal{R}$) is defined as:
\begin{equation}
\mathcal{R} = \frac{10\GeV}{\min(\max(\pt,50\GeV),200\GeV)},
\end{equation}
where the lepton \pt is measured in \GeV.
Corrections are applied to the computation of the lepton isolation in order to account for the effect of pileup using the effective area method~\cite{FastJet1}.
Two categories of leptons are defined, a ``tight'' lepton, which has $I<0.1$ and also passes the relevant identification criteria above,  and a ``loose'' lepton, which has $I < 0.4$.
In addition, the definition of ``loose'' electrons includes a relaxed requirement on the MVA discriminant, and ``loose'' muons have relaxed requirements on several of the aforementioned identification requirements.
The signal efficiencies for ``tight'' and ``loose'' electrons (muons) are $\approx$88\% ($\approx$97\%) and $\approx$95\% ($\approx$100\%) for $\abs{\eta}<2.5$ (2.4), respectively, excluding the barrel-endcap transition region ($1.44<\abs{\eta}<1.57$) for electrons.

Data-to-simulation scale factors to correct for imperfect detector simulation are obtained using the ``tag-and-probe'' method~\cite{CMS:2011aa} for lepton trigger, identification, and isolation, as functions of the lepton \pt and $\eta$.

Jets are clustered from the reconstructed PF candidates using the anti-\kt algorithm~\cite{Cacciari:2008gp} implemented in the \textsc{FastJet} package~\cite{FastJet1,FastJet2,FastJet3} with a distance
parameter of $0.4$ (AK4) and are required to satisfy $\pt>30$\GeV and $\abs{\eta}<2.4$.
Additional selection criteria are applied to remove spurious energy deposits originating from isolated
noise patterns in certain HCAL regions and from anomalous signals caused by particles depositing
energy in the silicon avalanche photodiodes used in the ECAL barrel region~\cite{CMS-PAS-JME-16-003}.
Jets that overlap with leptons have the four-momentum of any shared lepton subtracted from the jet four-momentum.
Jet energy corrections are applied for residual nonuniformity, nonlinearity of
the detector response, and the level of pileup in the event~\cite{JECold,JEC}.

In the single-lepton final state analysis, jets are tagged as originating from the decay of a bottom quark using a combined secondary vertex (CSVv2) algorithm~\cite{btagRun2}, which classifies jets based on the distance between their vertex and the primary vertex, along with observables such as track impact parameter. At the working point chosen, the efficiency for correctly tagging jets from bottom quark decays is between 40--65\%, depending on the jet \pt. The efficiency of tagging charm hadron jets is approximately 12\%, averaged over jet \pt, while the probability of mistagging light-flavor jets is roughly 1\%.

Large-radius jets are also reconstructed using the anti-\kt algorithm, with a distance parameter of 0.8 (AK8), and are used to tag hadronic decays of Lorentz-boosted top quarks or {\PW} bosons in the single-lepton final state analysis. Two variables are used to classify AK8 jets as originating from merged top quark decays ({\cPqt} tagging): the jet mass after grooming with the soft-drop algorithm~\cite{softdrop} and the ratio of $N$-subjettiness variables $\tau_{3}/\tau_{2}$~\cite{NSUBJETS}, a variable that provides strong discrimination between AK8 jets with two and three subjets.
For an AK8 jet to be labeled as {\cPqt} tagged, it must have $\pt>400\GeV$, soft-drop mass between 105 and 220\GeV, and the ratio $\tau_{3}/\tau_{2}$ less than 0.81.
This set of {\cPqt} tagging requirements yields an efficiency of roughly 60\% and a mistag rate of roughly 3\% for the \pt range considered.
Data-to-simulation scale factors~\cite{CMS-PAS-JME-16-003} are applied to events containing {\cPqt}-tagged jets in order to match the performance in the simulation to that seen in data.

If an AK8 jet fails the top quark identification criteria, it is considered for classification as a merged hadronic {\PW} boson decay ({\PW} tagging).
An AK8 jet is labeled as {\PW} tagged if it has $\pt>200\GeV$, pruned mass between 65 and 105\GeV, and a ratio of $N$-subjettiness variables $\tau_{2}/\tau_{1}$ smaller than 0.6, where the pruned mass is the mass of the jet after removing the soft and wide-angle radiated partons~\cite{pruning}.
This set of requirements used to select {\PW}-tagged jets yields a signal efficiency of 60--80\% and a mistag rate of 20--5\%, depending on the \pt of the AK8 jet.
The pruned mass scale is found to be consistent between data and simulation, but the mass resolution is found to be better in simulation and hence it is smeared in simulated events to match the resolution seen in data.
Data-to-simulation scale factors~\cite{CMS-PAS-JME-16-003} are also applied in order to match the performance of the {\PW} tagging in simulation to that seen in data.

The missing transverse momentum (\ptvecmiss) is defined as the negative of the vector \pt sum of all reconstructed PF candidates in an event and its magnitude is denoted as \ptmiss.
Energy scale corrections applied to jets are also propagated to \ptmiss.

\section{Same-sign dilepton final state}\label{sec:SSDL}

The search in the same-sign dilepton final state takes advantage of the rare signature of same-sign leptons, as well as the significant number of other high-\pt leptons and jets from the decay of the other \xft particle in the event.

The background contributions associated with this channel fall into three main categories: same-sign prompt leptons (SSP), opposite-sign prompt leptons (ChargeMisID), and same-sign nonprompt dilepton (Nonprompt).
The SSP background consists of SM processes that give prompt, same-sign dilepton signatures, where a prompt lepton is defined as one originating from the direct decay of either a {\PW} or {\PZ} boson. The contribution of these processes to the signal region is estimated using simulation.
The ChargeMisID background is composed of events that contain two opposite-sign leptons, but have the charge of one lepton mismeasured. This contribution is estimated from data.
The Nonprompt background consists of events that contain at least one nonprompt lepton passing the lepton selection criteria. Such events arise from jets misidentified as leptons, nonprompt leptons from heavy-flavor decays or conversions in the detector material, etc. This contribution is also estimated using control samples in data.

We first require two same-sign leptons that pass the tight definition given in Section~\ref{sec:Reconstruction}. The same-sign lepton pair that maximizes the scalar \pt sum of its constituents is taken as the signal pair.
Because the same-sign dilepton final state sample was collected in two different triggering eras, different \pt requirements are placed on the pair according to the triggering era in order to ensure that the trigger has reached full efficiency.
For the early\,(late) triggering era, the leading lepton is required to have $\pt > 40$\,(40)\GeV while the subleading lepton is required to have $\pt>35$\,(30)\GeV.

A set of preselection requirements is defined as follows.
First, the invariant mass of the same-sign lepton pair is required to be greater than 20\GeV (quarkonia veto) and the event is required to contain at least two AK4 jets passing the requirements outlined above.
Second, events containing a {\PZ} boson are removed by vetoing any event with an opposite-sign, same-flavor pair of leptons having an invariant mass within 15\GeV of the mass of the {\PZ} boson.
For the dielectron channel, this requirement is extended to the pair of same-sign electrons as well, in order to veto ChargeMisID background events. This eliminates the majority of Drell--Yan (DY) events, which would otherwise be a major contributor to the ChargeMisID background, without adversely affecting our signal efficiency.

After the preselection, two analysis-specific variables are defined as follows. The number of constituents (\nconst) is the number of AK4 jets in the event together with the number of additional (\ie not in the same-sign pair) leptons passing the tight definition. The \HTl variable is the scalar \pt sum of all constituents including the same-sign pair.

\begin{figure}[hbtp]
\centering
\includegraphics[width=0.49\textwidth]{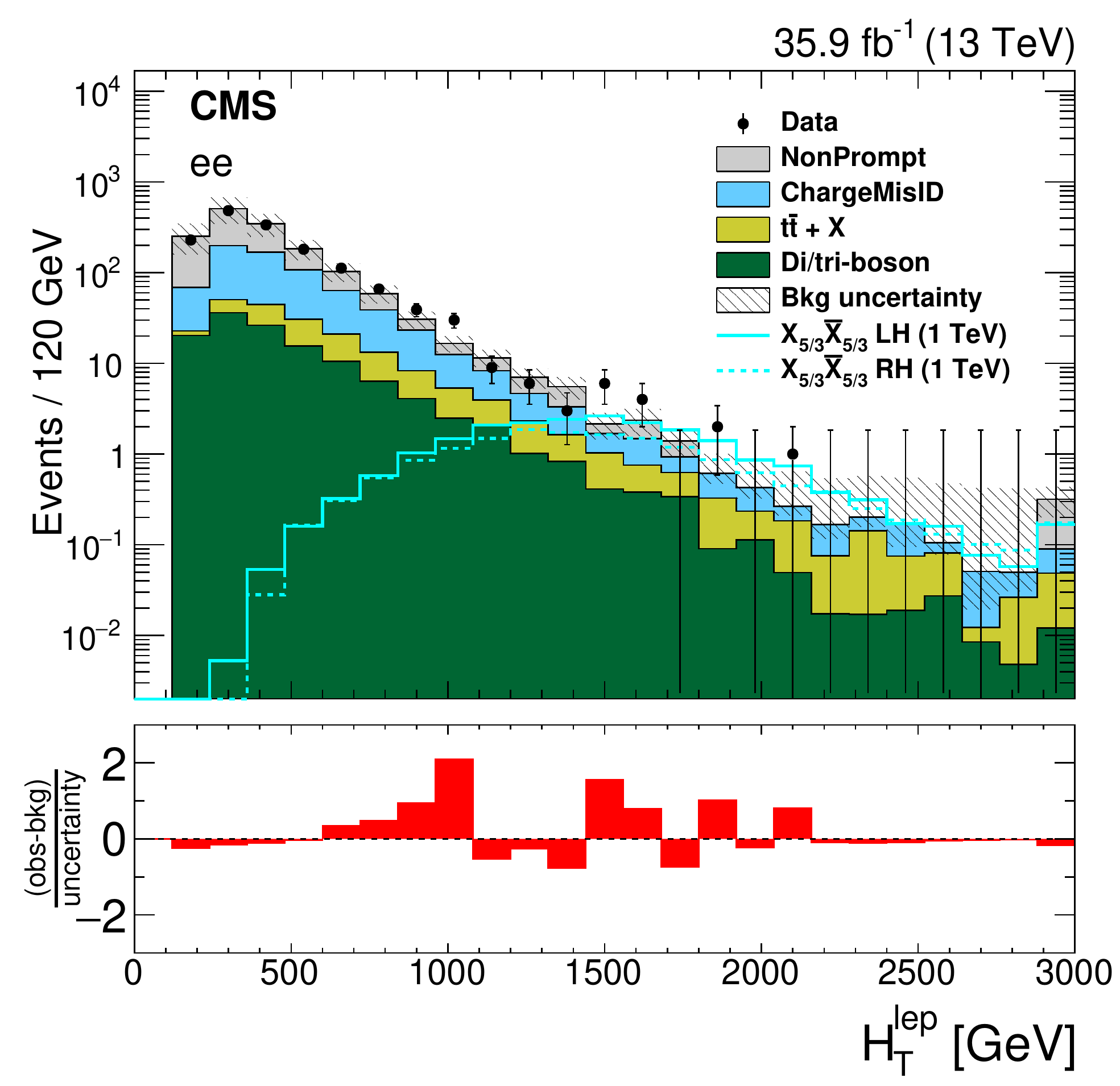}
\includegraphics[width=0.49\textwidth]{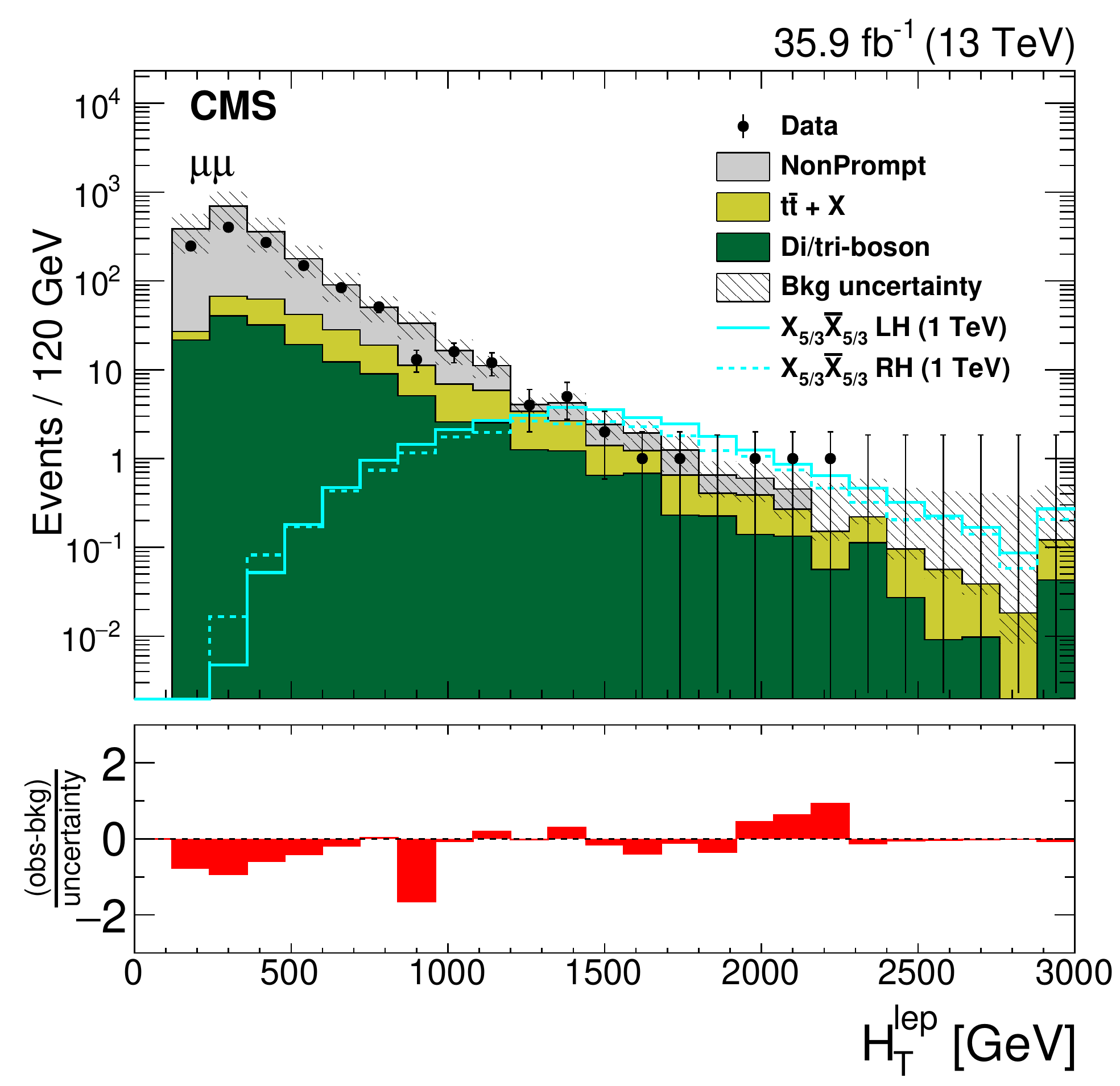}
\includegraphics[width=0.49\textwidth]{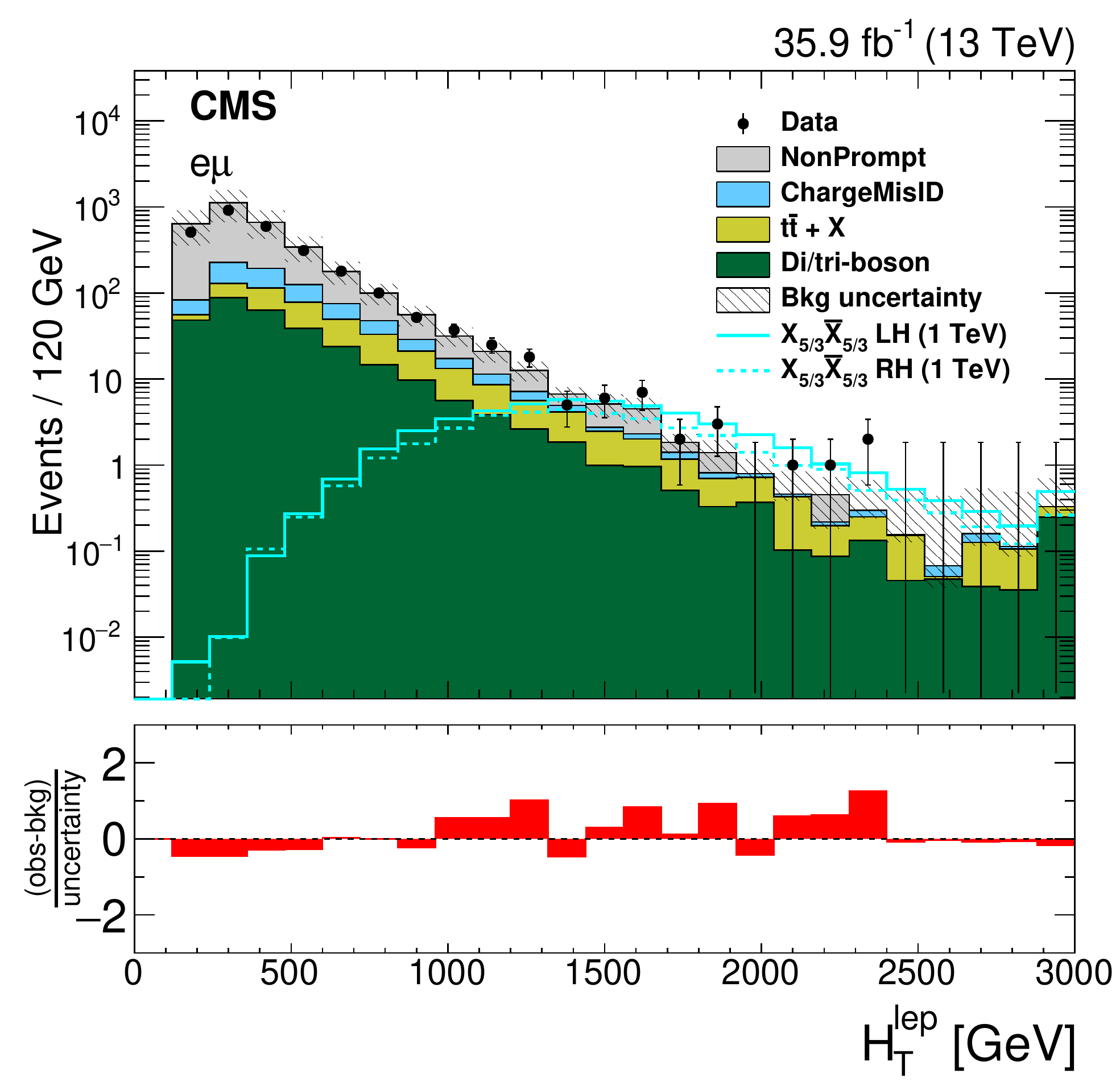}
\includegraphics[width=0.49\textwidth]{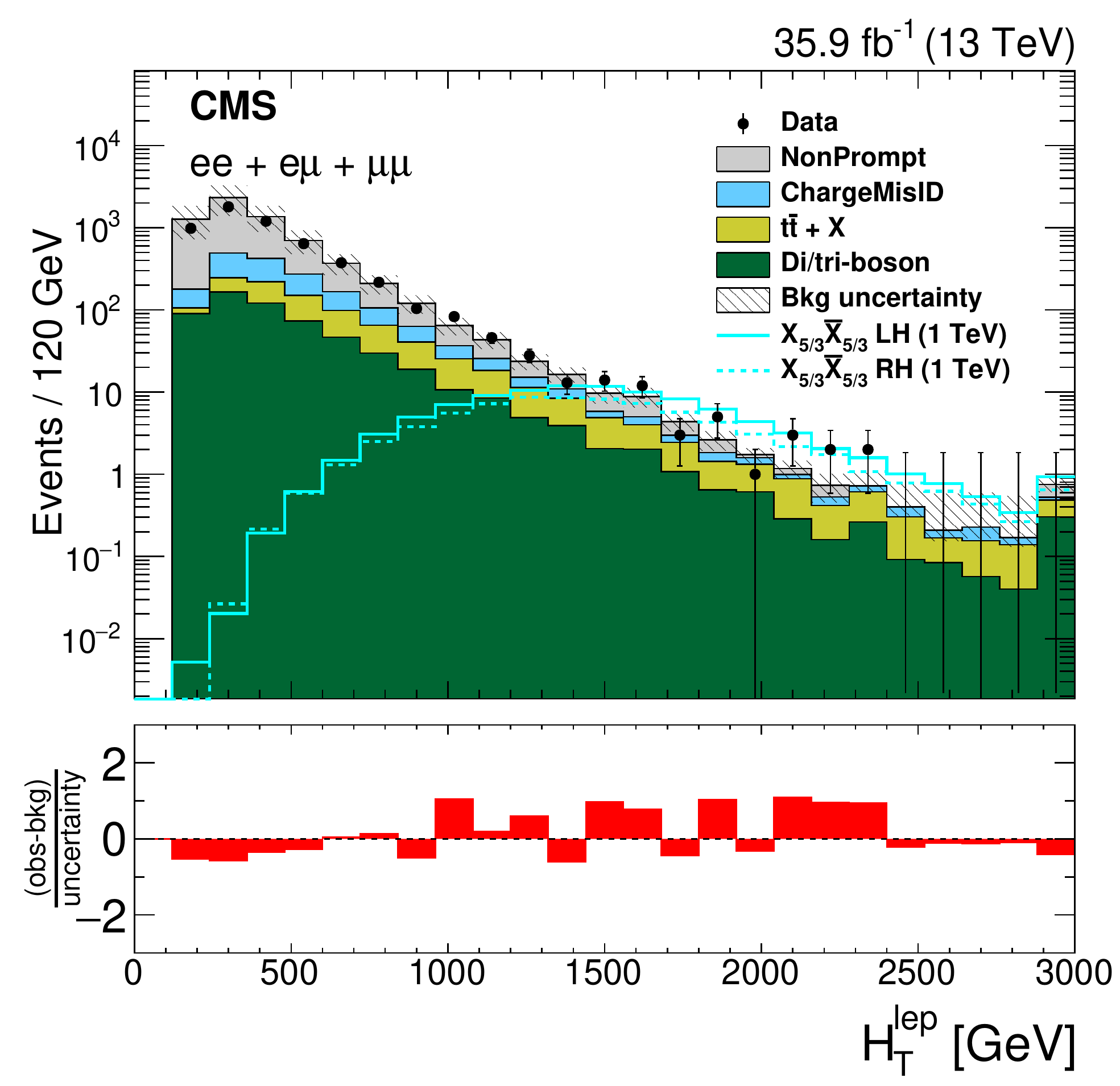}
\caption{The $\HTl$ distributions after the same-sign dilepton requirement, {\PZ} boson and quarkonia lepton invariant mass vetoes, and the requirement of at least two AK4 jets in the event, for dielectron (upper left), dimuon (upper right), electron-muon (lower left) final states, and their combination (lower right). The hatched area shows the combined systematic and statistical uncertainty in the background prediction for each bin.
The last bin includes overflow events.
The lower panel in each plot shows the difference between the observed and the predicted numbers of events divided by the total uncertainty. The total uncertainty is calculated as the sum in quadrature of the statistical uncertainty in the observed measurement and the uncertainty in the background, including both statistical and systematic components. Also shown are the expected signal distributions for a 1\TeV \xft with LH (solid line) and RH (dashed line) couplings.}
\label{fig:AK4HT}
\end{figure}

The criteria on these two variables are optimized for expected signal significance and the final requirements are $\nconst\geq5$ and $\HTl>1200\GeV$.
Figure~\ref{fig:AK4HT} shows the $\HTl$ distributions at the preselection level; the distributions of the {\nconst} variable (not shown) were also confirmed to be well described.

\subsection{Background modeling}

In this section, we summarize the background modeling used in the same-sign dilepton search.
The estimated contribution for all backgrounds is presented in Table~\ref{tab:SummaryYields}.
For additional details see Ref.~\cite{B2G-15-006}.

\subsubsection{Same-sign prompt lepton background}

The SSP background consists of processes with multiple {\PW} or {\PZ} bosons decaying to leptons, the bosons themselves either being created directly or through the decay of a top quark.
The contributions from these processes are estimated using the simulation as described in Section~\ref{sec:Simulation}.
The systematic uncertainties included for the SSP background are discussed in Section~\ref{sec:Systematics}.

\subsubsection{Opposite-sign prompt lepton background}\label{sec:chargemisid}

Background events in the ChargeMisID category arise from a pair of opposite-sign prompt leptons where the charge of one lepton is mismeasured, yielding a pair of same-sign leptons.
The charge misidentification probability for muons is much smaller and hence is considered negligible~\cite{susy2015}.
For electrons, the probability of charge misidentification is measured using observed DY events by requiring a pair of electrons with an invariant mass (driven by ECAL information) between 81 and 111\GeV.
The charge misidentification probability is binned by $\abs{\eta}$ of the electron, and split into three different \pt regions: below 100\GeV, between 100 and 200\GeV, and above 200\GeV.
These regions capture the effects of the differences in charge consistency requirements for low- and high-\pt electrons, as well as any remaining inherent dependence of the charge misidentification probability on the electron \pt.
Values of the charge misidentification probability range from $10^{-4}$ for low-\pt electrons in the central part of the detector to a few percent for high-\pt electrons in the forward region of the detector.

To estimate the contribution of the ChargeMisID background, opposite-sign dilepton events that satisfy all signal region kinematic requirements are weighted by the relevant probability of charge misidentification according to the kinematics of the electron(s) in the opposite-sign pair.

To account for the differences seen in the overall charge misidentification rate between DY and \ttbar events in simulation (roughly 25\% higher in DY), and some small residual kinematic disagreements (\pt dependent variation of roughly 5\% or less), a 30\% systematic uncertainty is assigned to the estimate of the number of ChargeMisID background events.

\subsubsection{Same-sign nonprompt background}\label{sec:fakebkg}

The Nonprompt background arises from events where a nonprompt lepton (such as a lepton from a heavy-flavor decay, photon conversion, or a misidentified jet) passes the tight lepton identification requirements.
Contributions from these types of events are estimated using the ``Tight-Loose'' method as described in Ref.~\cite{susy2011}.
This method relies on collecting a sample of dilepton events where the leptons are allowed to pass the loose definition described previously, and then scaling those events by weights involving the probability of a loose prompt lepton to pass the tight definition (``prompt rate'') and the probability of a loose nonprompt lepton to pass the tight definition (``misidentification rate'').

The prompt rate is determined using the ``tag-and-probe'' technique with DY-enriched dilepton data where the invariant mass of the leptons is within 10\GeV of the {\PZ} boson mass.
For muons, the prompt rate is found to be flat to within a few percent as functions of $\eta$ and \pt and hence the average of 0.94 is taken.
The prompt rate for electrons is found to be flat versus $\eta$, but has a \pt dependence, which is taken into account and gives values for the prompt rate ranging from 0.80 to 0.95.

The misidentification rate is determined using a sample enriched in QCD multijet events.
The selection of this sample follows the approach described in Ref.~\cite{B2G-15-006} and requires exactly one loose lepton, at least one jet, low \ptmiss, and low $M_T$, where $M_T$ is the transverse mass of the lepton and \ptmiss. 
We also reject events if the invariant mass of the lepton and any jet is within 10\GeV of the {\PZ} boson mass.

Because of the significantly larger integrated luminosity used in this analysis, binning of the variation in the misidentification rate as a function of lepton $\eta$ is possible; the values obtained range from 0.16 to 0.25\,(0.34) for electrons\,(muons), with the lower values corresponding to leptons in the central part of the detector.

The uncertainty in the estimation of the Nonprompt background is derived by comparing the variation between the misidentification rates measured from different types of nonprompt lepton candidates, categorized by the generator-level origin of the nonprompt lepton; the variation in kinematic dependence of these misidentification rates with respect to \pt and $\eta$; and the overall level of closure seen in the method.
The above checks are all performed using \ttbar MC events.
To ensure that all effects are covered, a 50\% uncertainty is assigned to the estimate of the Nonprompt background.

\subsection{Event yields}

Summing over the three dilepton final states, between 1.8 (2.4) and 3.4 (4.1)\% of the produced \xft pairs are expected to pass the full selection criteria for an LH (RH) signal, depending on the \xft mass.
The number of observed events, along with the expected number of background events broken down by category, is shown in Table~\ref{tab:SummaryYields}.
The background predictions in the table are derived after a ``background-only'' fit to the data as described in Section~\ref{sec:Results}, where the signal strength is assumed to be zero.
The fit increases the predicted Nonprompt background by less than its originally assigned uncertainty, and reduces the uncertainty associated with this background by about 30\%.
Also shown is the number of expected signal events for an RH \xft with mass 1\TeV.
The observed number of events in the signal region categories are compatible with the background predictions.

\begin{table}[ht]
  \centering
  \topcaption{Summary of yields from simulated prompt same-sign dilepton (SSP MC), same-sign nonprompt (Nonprompt), and opposite-sign prompt (ChargeMisID) backgrounds after the full analysis selection. Also shown are the number of expected events for an RH \xft particle with a mass of 1\TeV. The uncertainties include both statistical and all systematic components (as described in Section~\ref{sec:Systematics}). The number of events and uncertainties correspond to the background-only fit to data for the background, while for the signal they are based on the yields before the fit to data.}
  \cmsTable{
	  \begin{tabular}{l c r@{\,$\pm$\,}l r@{\,$\pm$\,}l c c c}
      Channel       & RH \xft (1\TeV{}) & \multicolumn{2}{c}{SSP MC}  & \multicolumn{2}{c}{Nonprompt} & ChargeMisID & Total bkg.   & Data \\
      \hline
      Dielectron    & $11.6\pm0.8$    & 3.9&0.3  & 4.6&1.7  & $2.4\pm0.7$ & $10.9\pm1.9$ & 10   \\
      Dimuon        & $16.1\pm1.2$    & 5.7&0.5  & 5.5&1.9  & \NA         & $11.2\pm2.0$ & 12   \\
      Electron-muon & $26.9\pm1.9$    & 10.3&0.8 & 11.3&3.6 & $1.7\pm0.5$ & $23.2\pm3.7$ & 26   \\
      \end{tabular}
  }
\label{tab:SummaryYields}
\end{table}

\section{Single-lepton final state}\label{sec:LJets}

The single-lepton final state targets events where one of the four {\PW} bosons in the event decays leptonically and the others decay hadronically (including hadronic tau decays).
Events are required to have exactly one tight lepton with $\pt >80\GeV$.
An event is discarded if it contains another lepton that passes the loose identification criteria and has $\pt>10\GeV$.
In order to limit the background contributions from QCD multijet events, selected events are required to have $\ptmiss > 100\GeV$ and the AK4 jet that is closest to the lepton is either required to be separated by $\Delta R > 0.4$, where $\Delta R = \sqrt{\smash[b]{(\Delta\eta)^2+(\Delta\phi)^2}}$, or the magnitude of the lepton momentum that is transverse to the jet axis is required to be greater than 40\GeV.

Since the signal topology includes significant levels of hadronic activity, events are also required to have at least four AK4 jets, and the leading and subleading jets are required to have a \pt greater than 450 and 150\GeV, respectively.
At least one of the four AK4 jets is required to pass the {\cPqb} tagging requirement.

\begin{figure}[hbtp]
\centering
\includegraphics[width=0.49\textwidth]{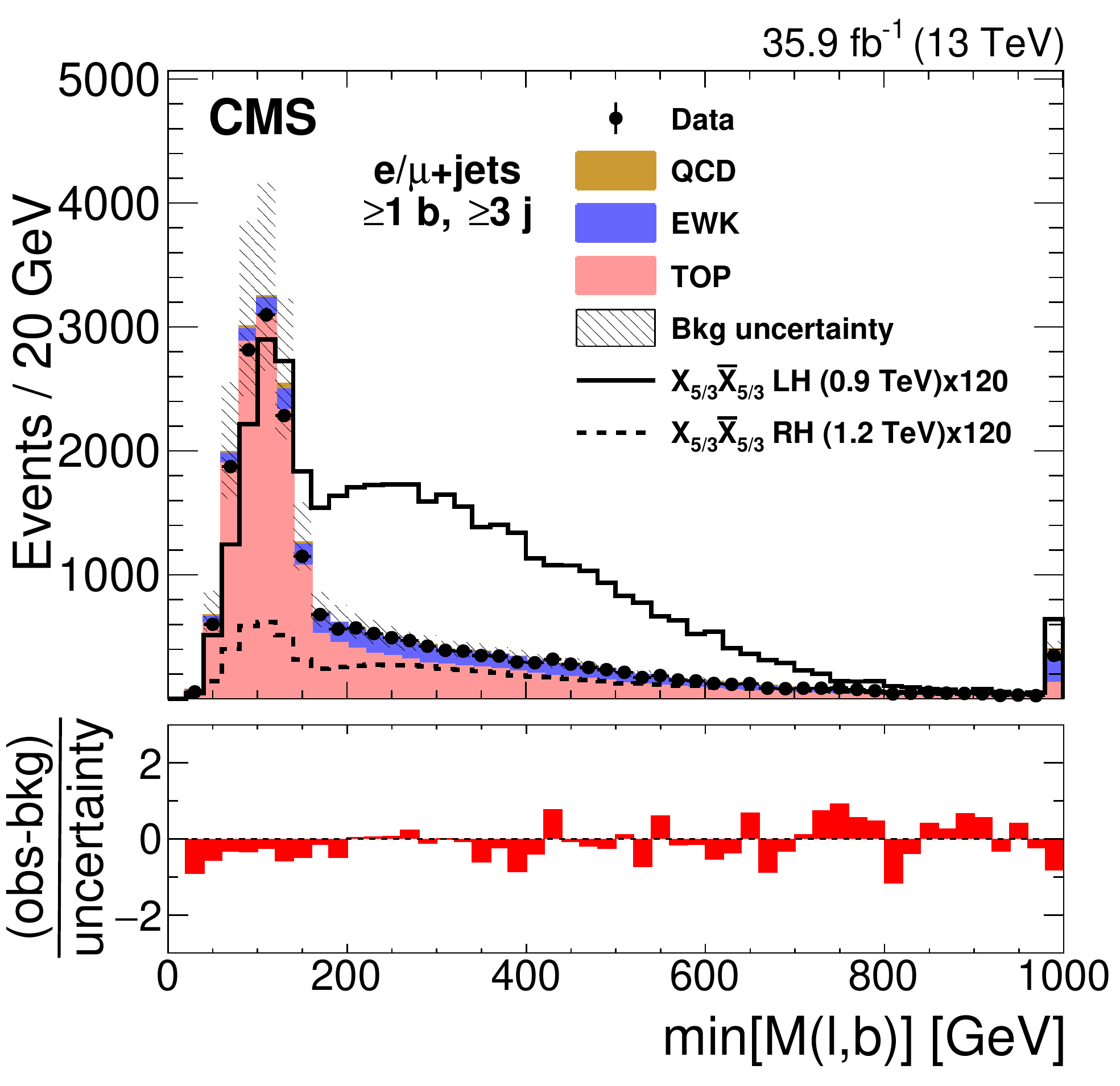}
\includegraphics[width=0.49\textwidth]{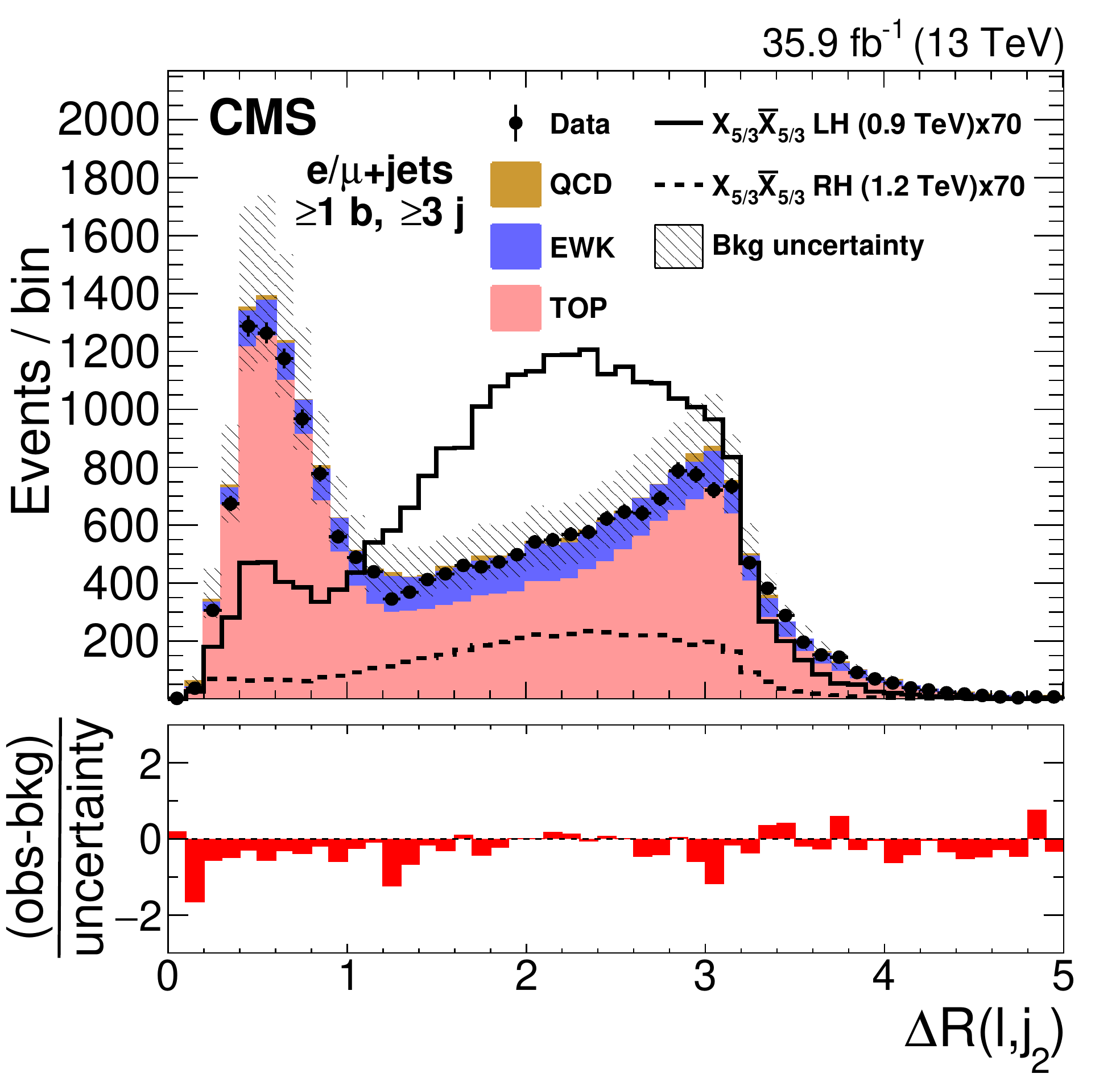}
\caption{Distributions of \minmlb (left) and \drlj (right) in data and simulation for events with at least three AK4 jets, including a leading (subleading) jet with $\pt > 250$ (150)\GeV, after combining the electron and muon channels.
Example signal distributions are also shown, scaled by a factor of 120\,(70) in the \minmlb\,(\drlj) distribution.
The last bin includes overflow events.
The lower panel in each plot shows the difference between the observed and the predicted numbers of events in that bin divided by the total uncertainty.
The total uncertainty is calculated as the sum in quadrature of the statistical uncertainty
in the observed measurement and the statistical and systematic uncertainties in the background.}
\label{fig:presel}
\end{figure}

Two observables are found to provide strong discrimination between signal and background events as in Ref.~\cite{B2G-15-006}: \drlj, the angular separation between the lepton and subleading AK4 jet, and \minmlb, the minimum mass reconstructed using the lepton and any AK4 jet in the event passing the {\cPqb} tagging requirement. Signal regions for this search are constructed from events with $\drlj > 1.0$, with the distribution of \minmlb used for signal extraction. Figure~\ref{fig:presel} shows the distributions for \drlj and \minmlb in events with at least three AK4 jets, including a leading (subleading) jet with $\pt > 250$ (150)\GeV prior to the fit to data.
The distribution of \minmlb for the background, dominated by \ttbar events,
features a sharp drop around 150\GeV, since, for such events,
this variable represents the visible mass of the top quark in the detector.
The \drlj variable shows that the subleading jets populate both the same and opposite hemisphere relative to the lepton in the background events,
whereas in the \xft signal events, the subleading jet is usually opposite to the lepton.

\subsection{Background modeling}

All of the background processes in the single-lepton analysis are modeled using the simulation.
In order to confirm that this modeling is correct, the agreement between simulation and data is checked for the dominant (\ttbar) and subdominant ({\PW}+jets) background processes using background-enriched control regions.
The control regions have the same conditions as the signal region, with the requirement on \drlj inverted such that $0.4<\drlj<1.0$.
The {\PW}+jets enriched control region also requires that no jet passes the {\cPqb} tagging requirements, and is split into categories of either zero or at least one {\PW}-tagged jet. The \ttbar enriched control region uses the {\cPqb} tagging requirements of the signal region and is split into either 1 or $\geq$2 {\cPqb}-tagged jet categories.
With the lack of {\cPqb}-tagged jets in the {\PW}+jets control region, the reconstructed mass of interest is modified to be the minimum mass of the lepton and any AK4 jet in the event (\minmlj).

\begin{figure}[hbtp]
\centering
\includegraphics[width=0.49\textwidth]{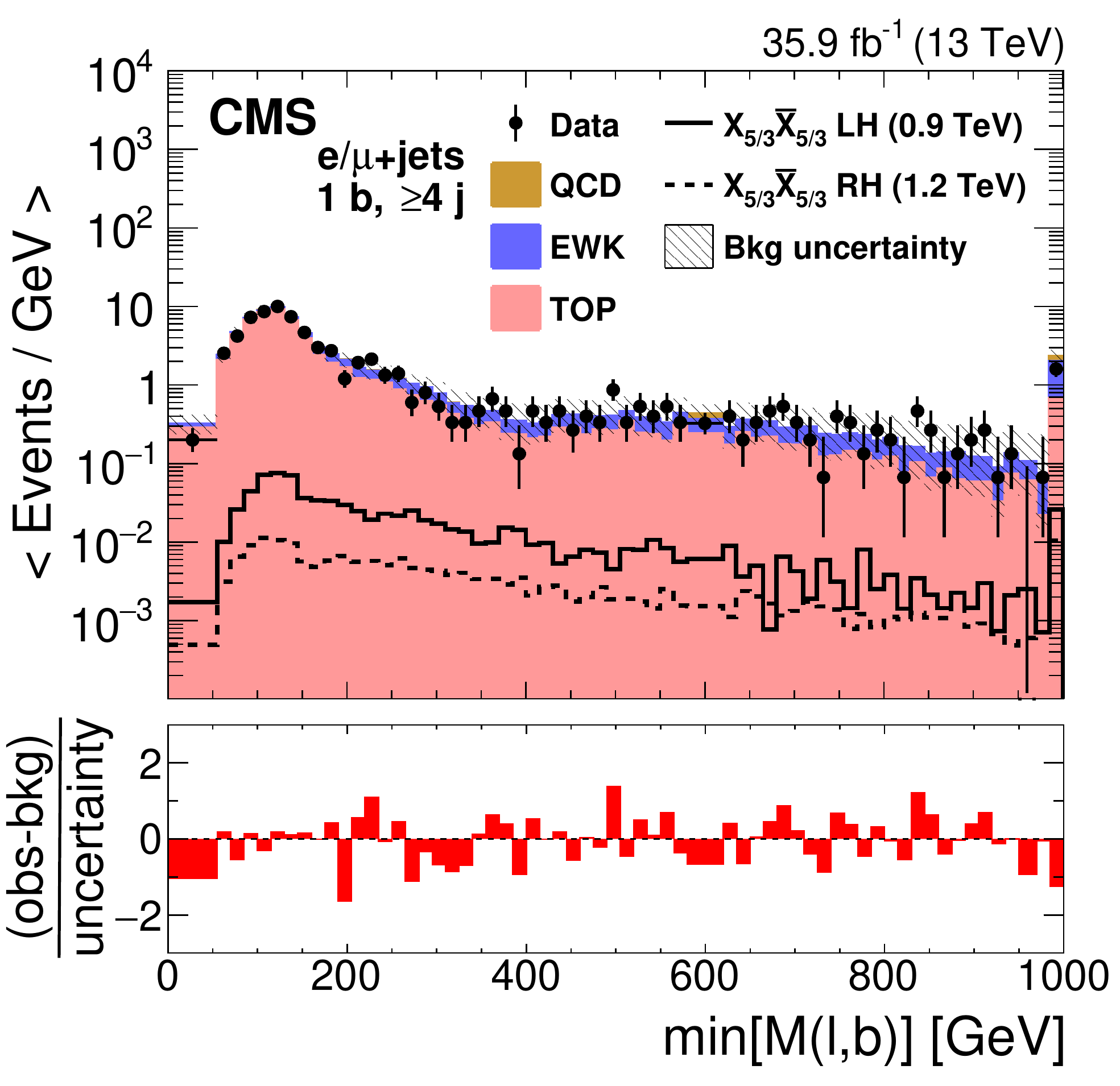}
\includegraphics[width=0.49\textwidth]{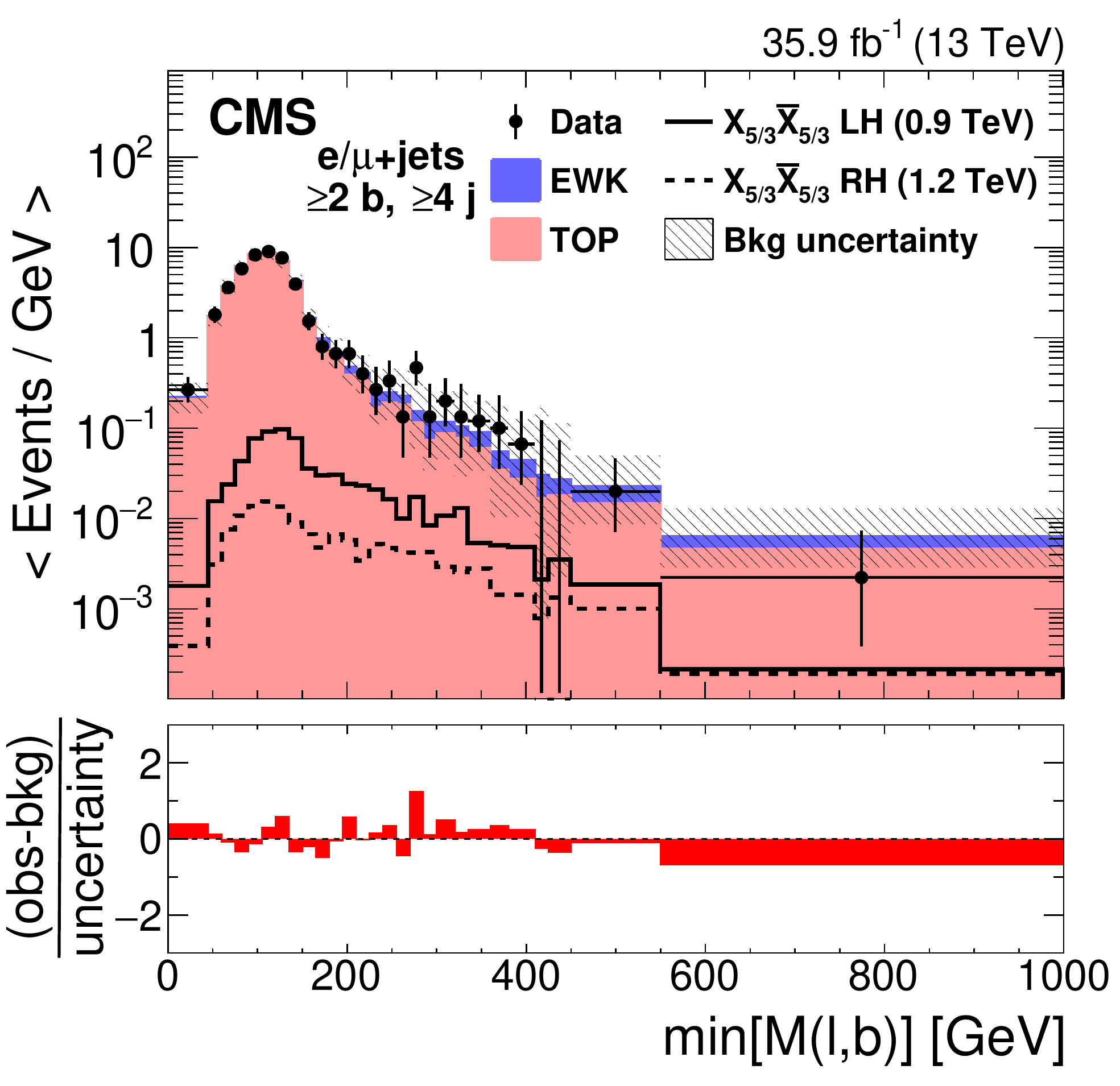}
\includegraphics[width=0.49\textwidth]{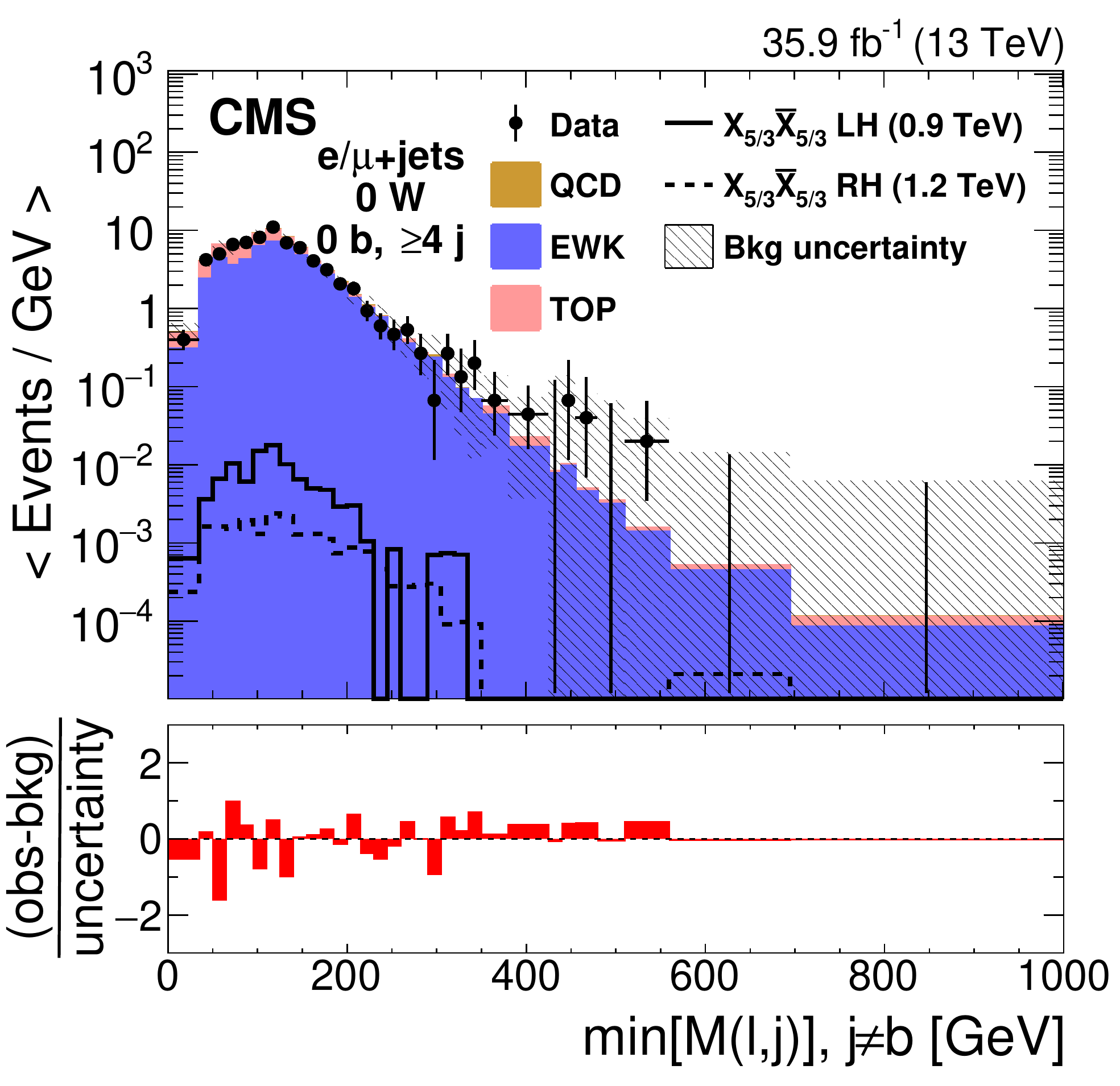}
\includegraphics[width=0.49\textwidth]{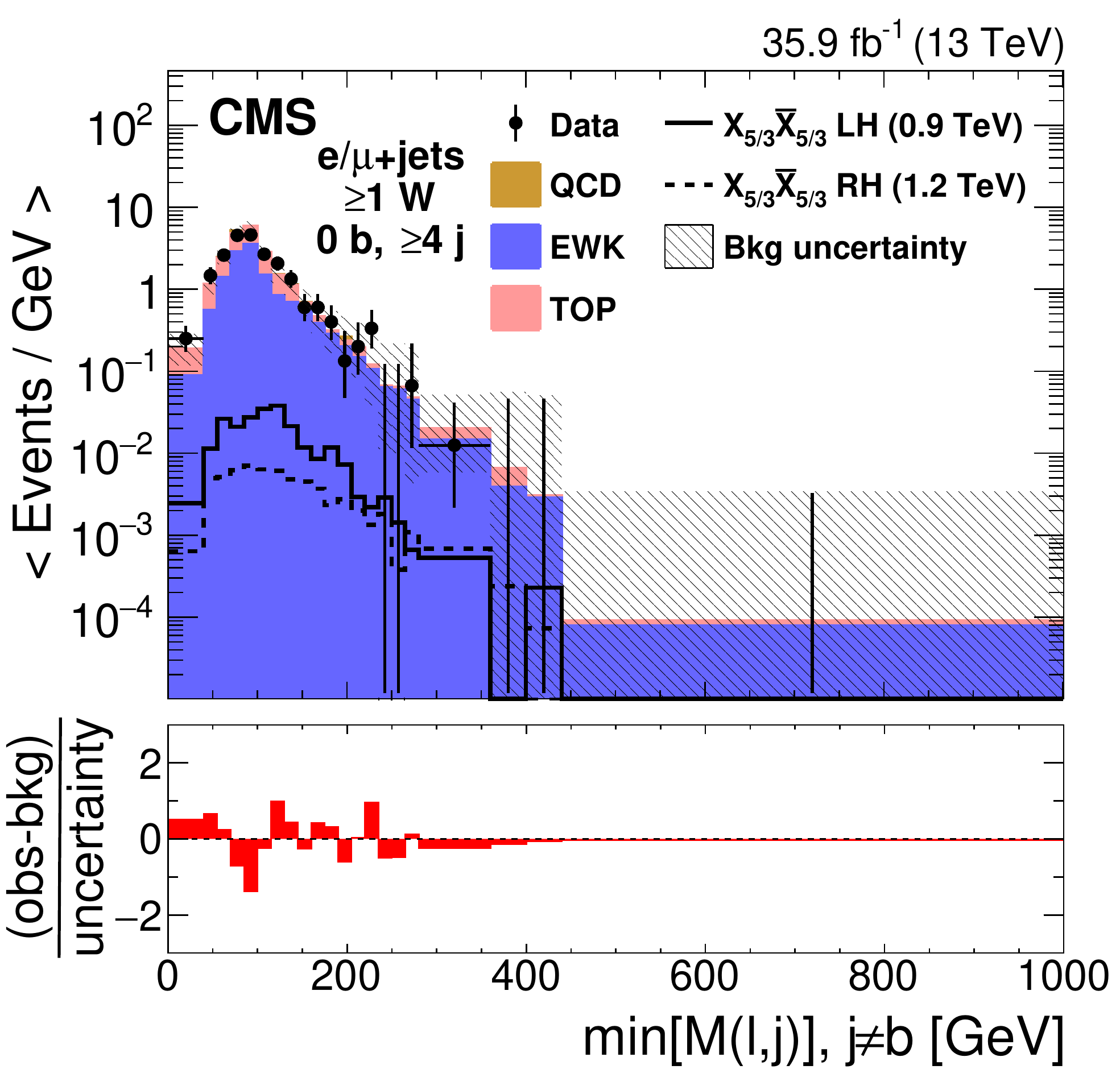}
\caption{Distributions of \minmlb in the \ttbar control region, for 1 {\cPqb}-tagged jet (upper left) and
$\ge$2 {\cPqb}-tagged jets (upper right) categories, and of \minmlj in the {\PW}+jets control region,
for 0 {\PW}-tagged jets (lower left) and  $\ge$1 {\PW}-tagged jets (lower right) categories.
Example signal distributions are also shown.
The background distributions correspond to background-only fit to data while signal distributions are before the fit to data.
Electron and muon event samples are combined.
The last bin includes overflow events and its content is divided by the bin width.
The distributions in each category have variable-size bins,
chosen so that the statistical uncertainty in the total background in each bin is less than 30\%.
The lower panel in each plot shows the difference between the observed and the predicted numbers of events
in that bin divided by the total uncertainty.
The total uncertainty is calculated as the sum in quadrature of the statistical uncertainty
in the observed measurement and the statistical and systematic uncertainties in the background-only fit to data.}
\label{fig:bkgCR}
\end{figure}

The agreement between the data and the SM prediction from simulation is checked in all control region categories and is found to be within the uncertainties in the prediction, which are detailed in Section~\ref{sec:Systematics}.
Figure~\ref{fig:bkgCR} shows the distributions of \minmlb and \minmlj for the \ttbar and {\PW}+jets enriched control regions, while Table~\ref{tab:CR} shows the predicted and observed numbers of events in each control region after the full analysis selection.
The background predictions in Fig.~\ref{fig:bkgCR} and Table~\ref{tab:CR} are given after the background-only fit to data using all categories in both final states, including both signal and control regions.

\begin{table}[htbp]
\centering
\topcaption{Expected (observed) numbers of background (data) events passing the final selection requirements,
in the \ttbar and {\PW}+jets control region ($0.4<\drlj<1.0$) categories, after combining the single-electron and single-muon channels.
The numbers of events expected from two example signals are also shown.
The event yields and their uncertainties correspond to the background-only fit to data for the background, while for the signal they are based on the values before the fit to data.}
\begin{tabular}{l cr@{\,$\pm$\,}l cr@{\,$\pm$\,}l r@{\,$\pm$\,}l cr@{\,$\pm$\,}l}
Sample
& \multicolumn{3}{c}{${\geq}0$ \cPqt, ${\geq}0$ \PW, 1 \cPqb}
& \multicolumn{3}{c}{${\geq}0$ \cPqt, ${\geq}0$ \PW, ${\geq}2$ \cPqb}
& \multicolumn{2}{c}{${\geq}0$ \cPqt, 0 \PW, 0 \cPqb}
& \multicolumn{3}{c}{${\geq}0$ \cPqt, ${\geq}1$ \PW, 0 \cPqb} \\
\hline
LH \xft (0.9\TeV)   && 13.15 & 0.61         && 10.90 & 0.58 & 1.46 & 0.27       && 3.60 & 0.36   \rule{0pt}{2.5ex}  \\
RH \xft (1.2\TeV)   && 3.02 & 0.13          && 2.34 & 0.12  & 0.32 & 0.06       && 1.00 & 0.08  \\
[\cmsTabSkip]
TOP                 && 953 & 97             && 668 & 72     & 274 & 30          && 134 & 14    \\
EWK                 && 200 & 16             && 29.5 & 3.1   & 789 & 57          && 204 & 15    \\
QCD                 && 12.9 & 5.4           && 1.05 & 0.55  & 14.5 & 4.6        && 7.2 & 3.9   \\
[\cmsTabSkip]
Total bkg.          && 1170 & 100           && 699 & 72     & 1077 & 70         && 345 & 23    \\
Data                & \multicolumn{3}{c}{1152} & \multicolumn{3}{c}{710} & \multicolumn{2}{c}{1062} & \multicolumn{3}{c}{335}  \\
\end{tabular}
\label{tab:CR}
\end{table}

\subsection{Event yields and template distributions}

\begin{figure}[hbtp]
\centering
\includegraphics[width=0.49\textwidth]{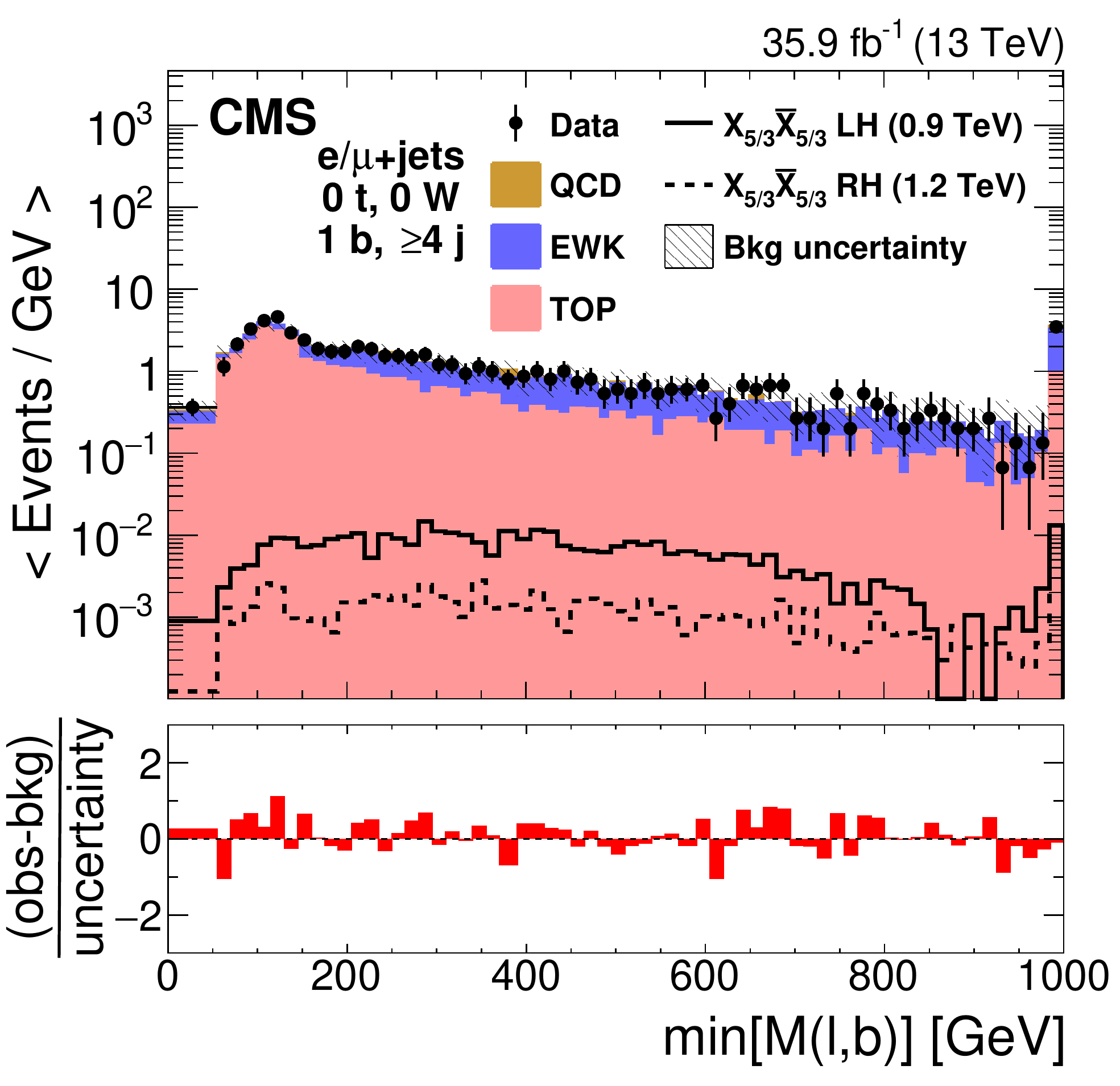}
\includegraphics[width=0.49\textwidth]{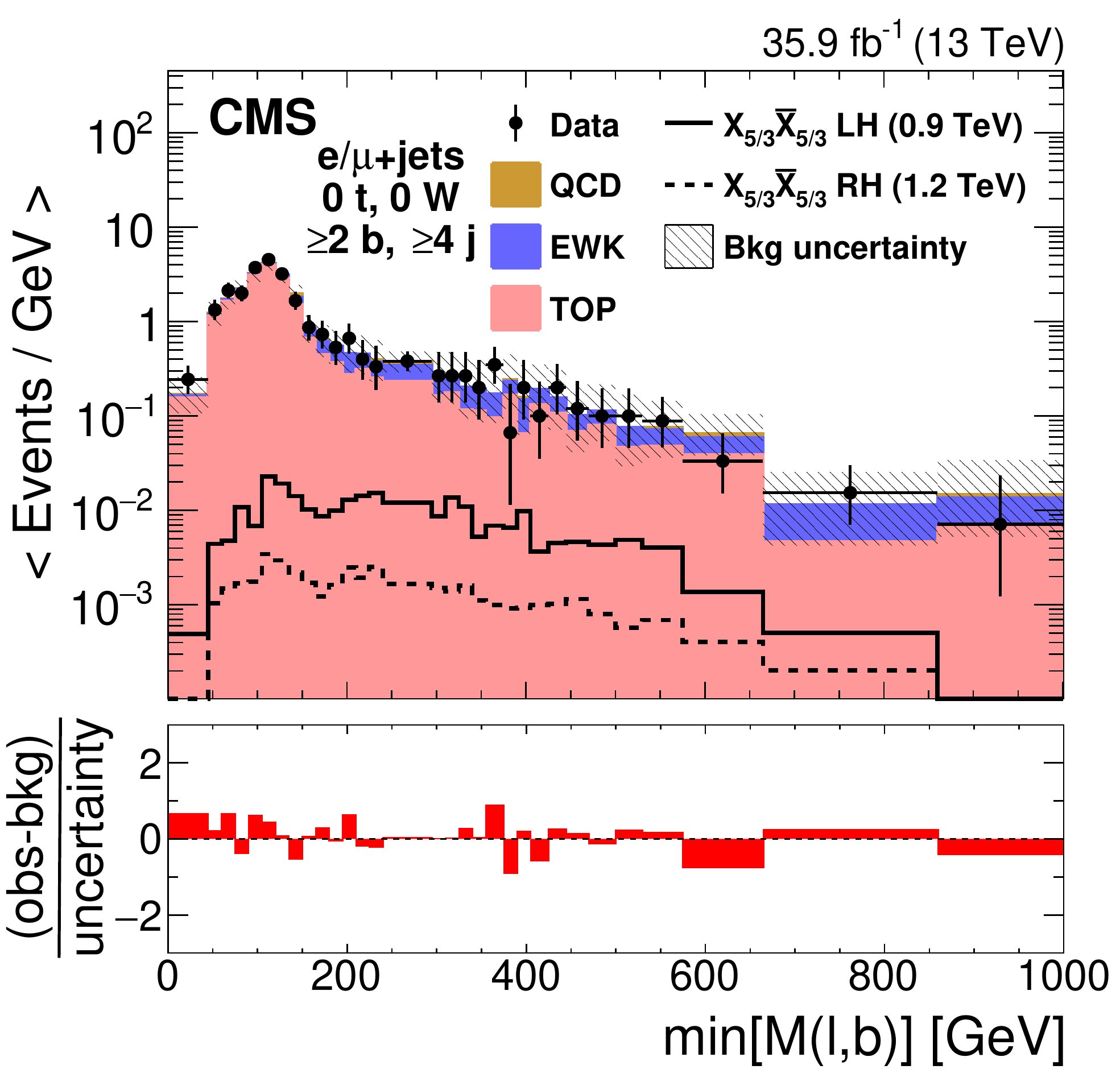}
\includegraphics[width=0.49\textwidth]{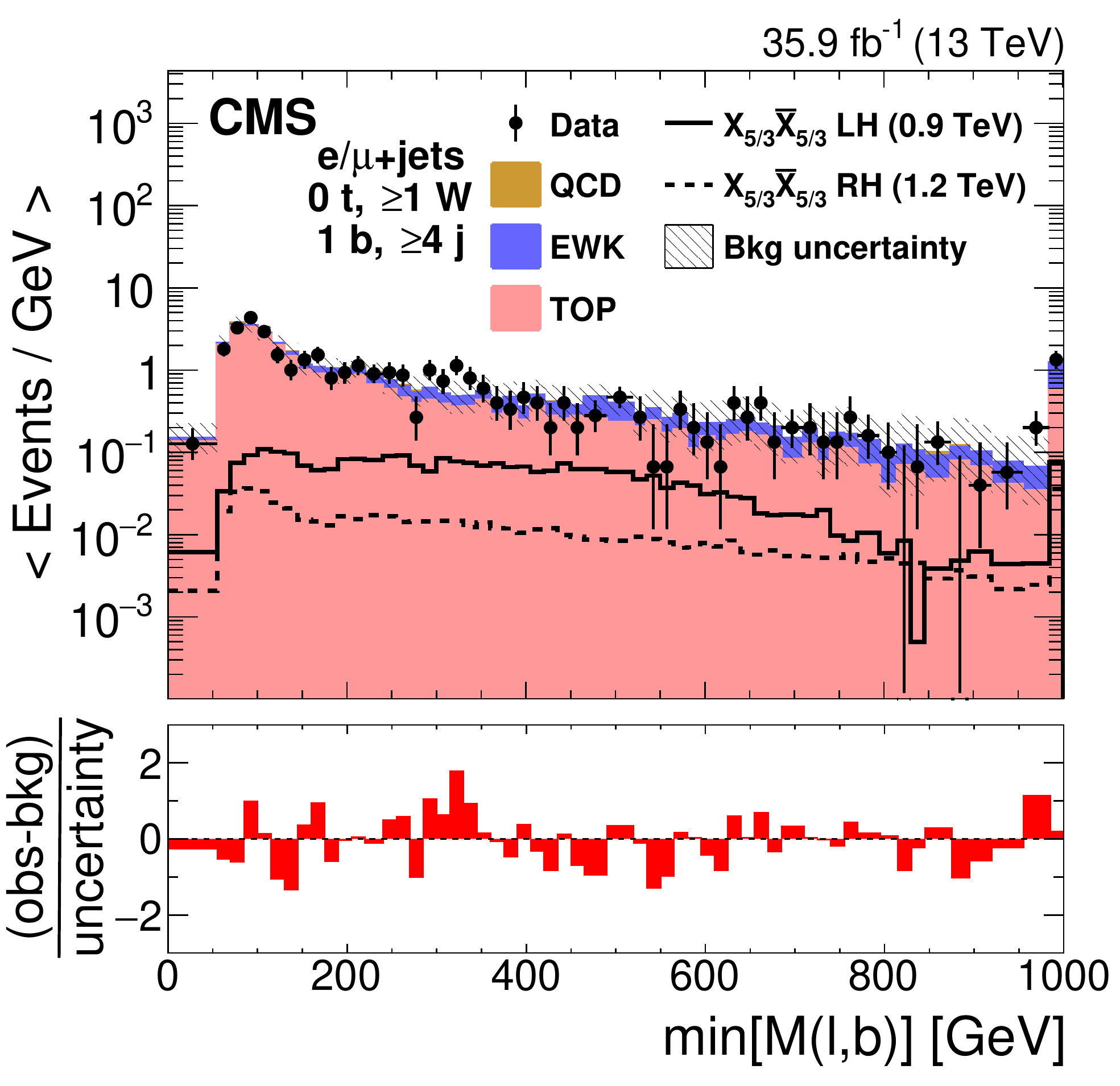}
\includegraphics[width=0.49\textwidth]{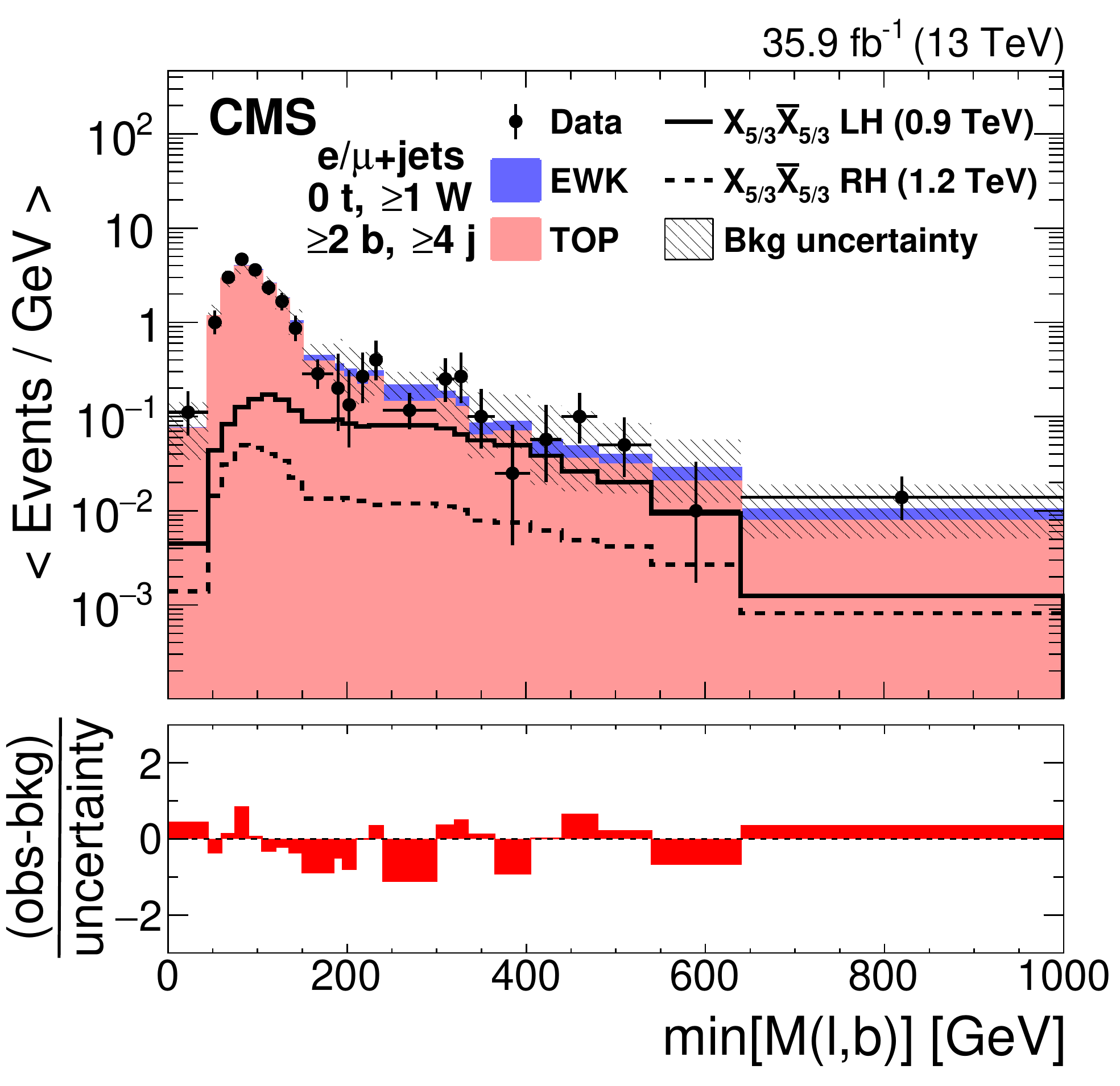}
\caption{Distributions of \minmlb in events with 0 {\cPqt}-tagged jets, 0 (upper) or $\geq$1 (lower) {\PW}-tagged jets,
and 1 (left) or $\geq$2 (right) {\cPqb}-tagged jets for the combined electron and muon samples in the signal region.
Example signal distributions are also shown.
The background distributions correspond to the background-only fit to data, while signal distributions are before the fit to data.
The last bin includes overflow events and its content is divided by the bin width.
The distributions in each category have variable-size bins,
chosen so that the statistical uncertainty in the total background in each bin is less than 30\%.
The lower panel in each plot shows the difference between the observed and the predicted numbers of events
in that bin divided by the total uncertainty.
The total uncertainty is calculated as the sum in quadrature of the statistical uncertainty
in the observed measurement and the statistical and systematic uncertainties in the background-only fit to data.}
\label{fig:Mlb0tcats}
\end{figure}

\begin{figure}[hbtp]
\centering
\includegraphics[width=0.49\textwidth]{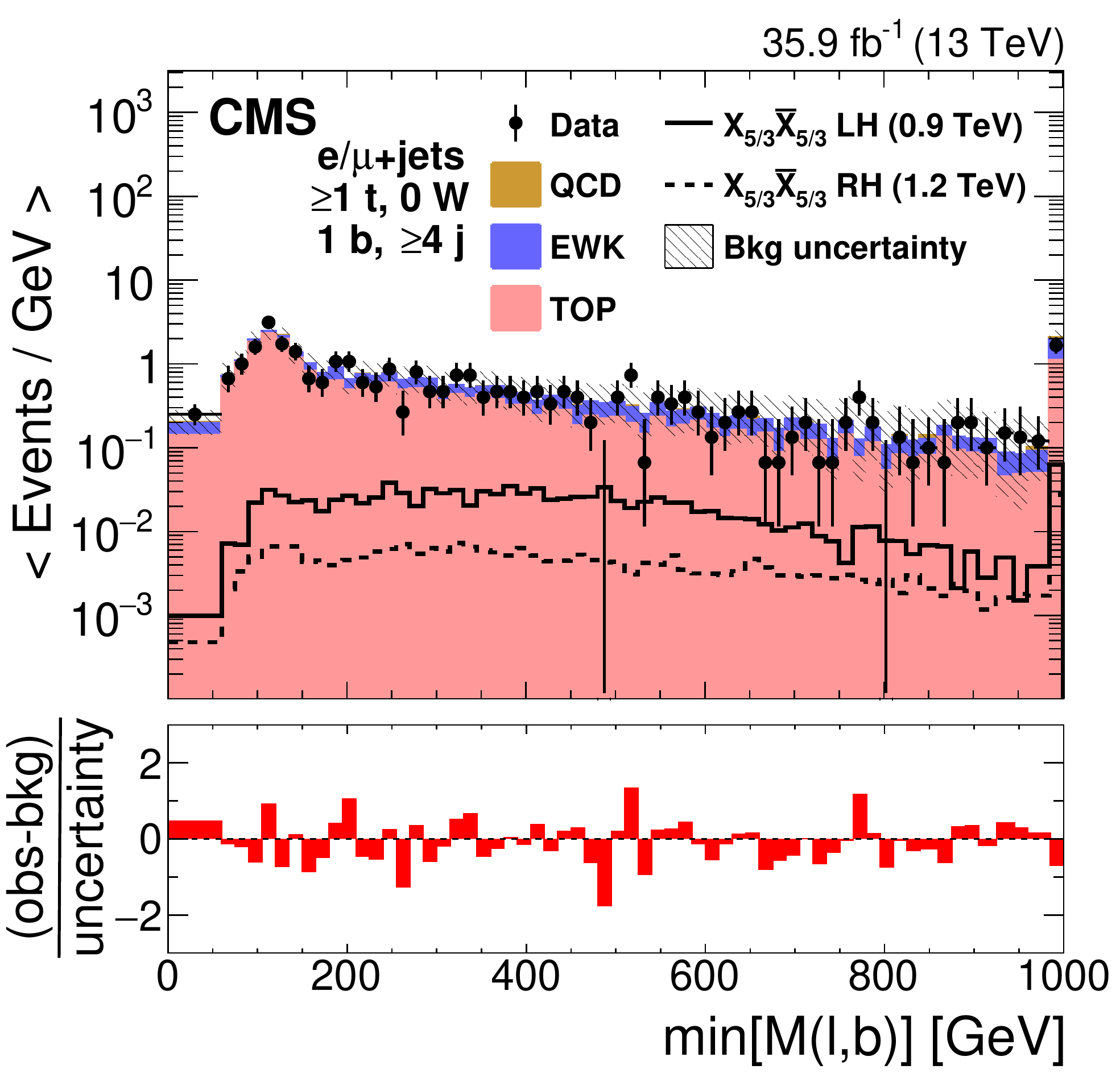}
\includegraphics[width=0.49\textwidth]{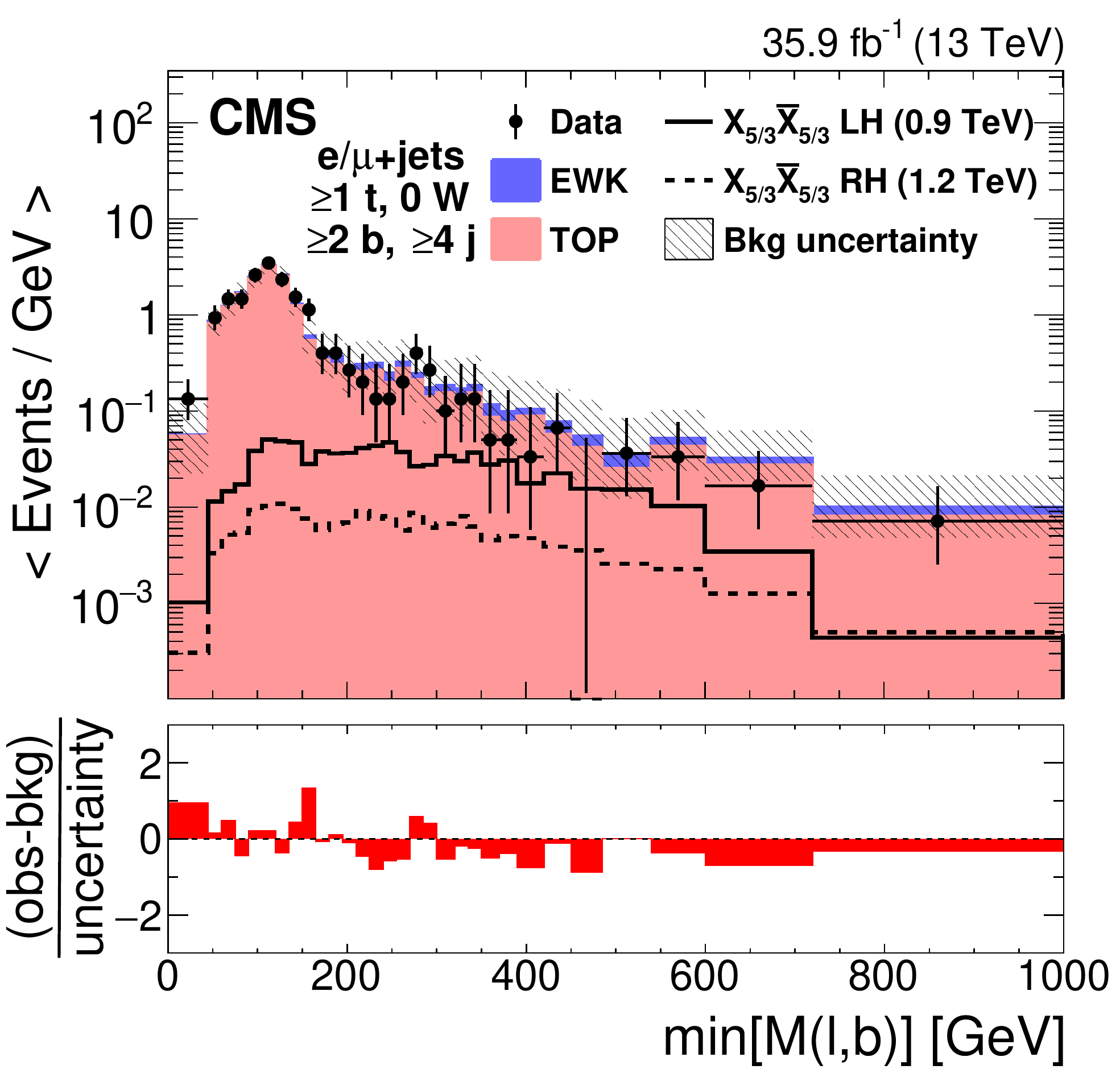}
\includegraphics[width=0.49\textwidth]{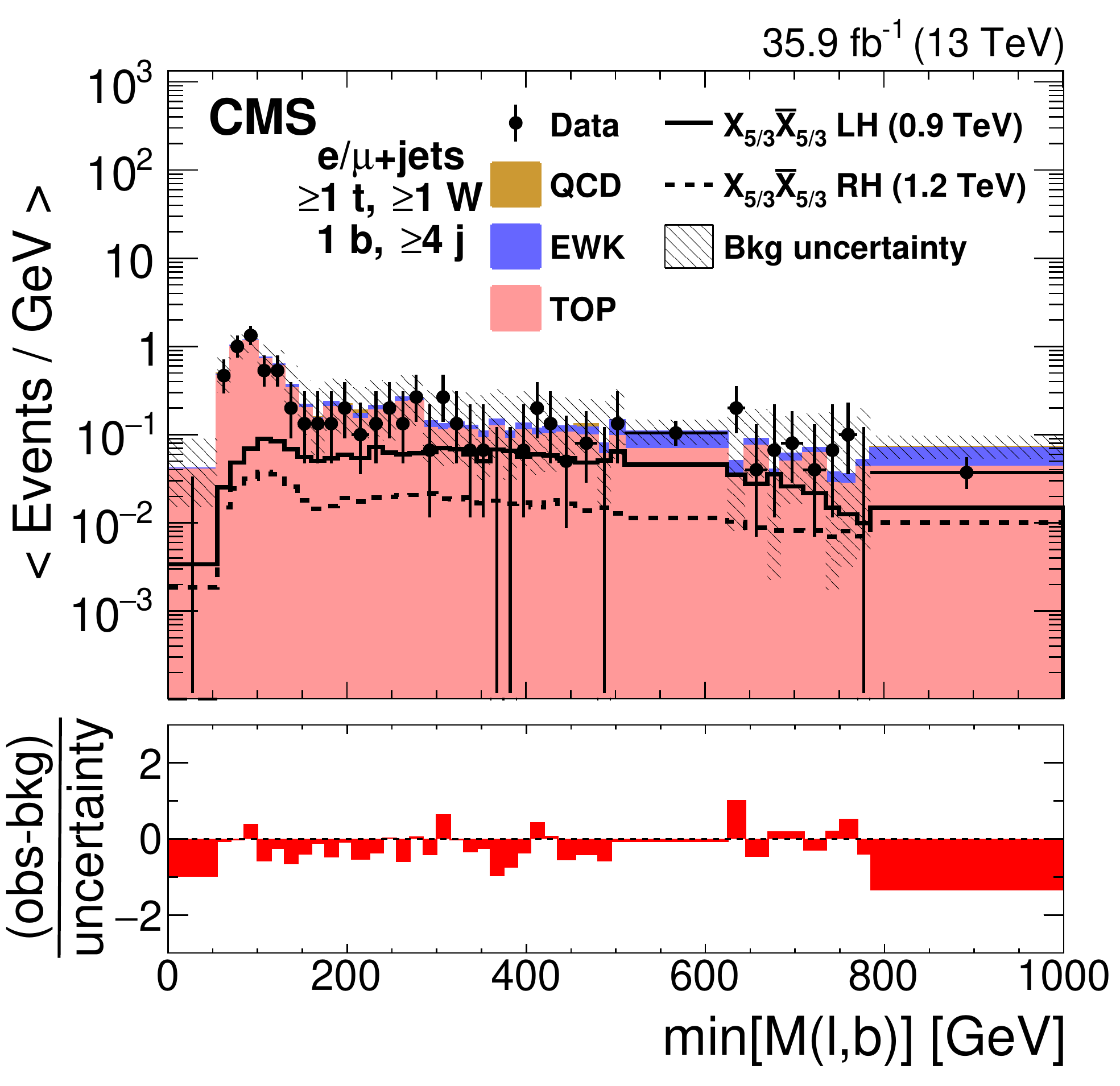}
\includegraphics[width=0.49\textwidth]{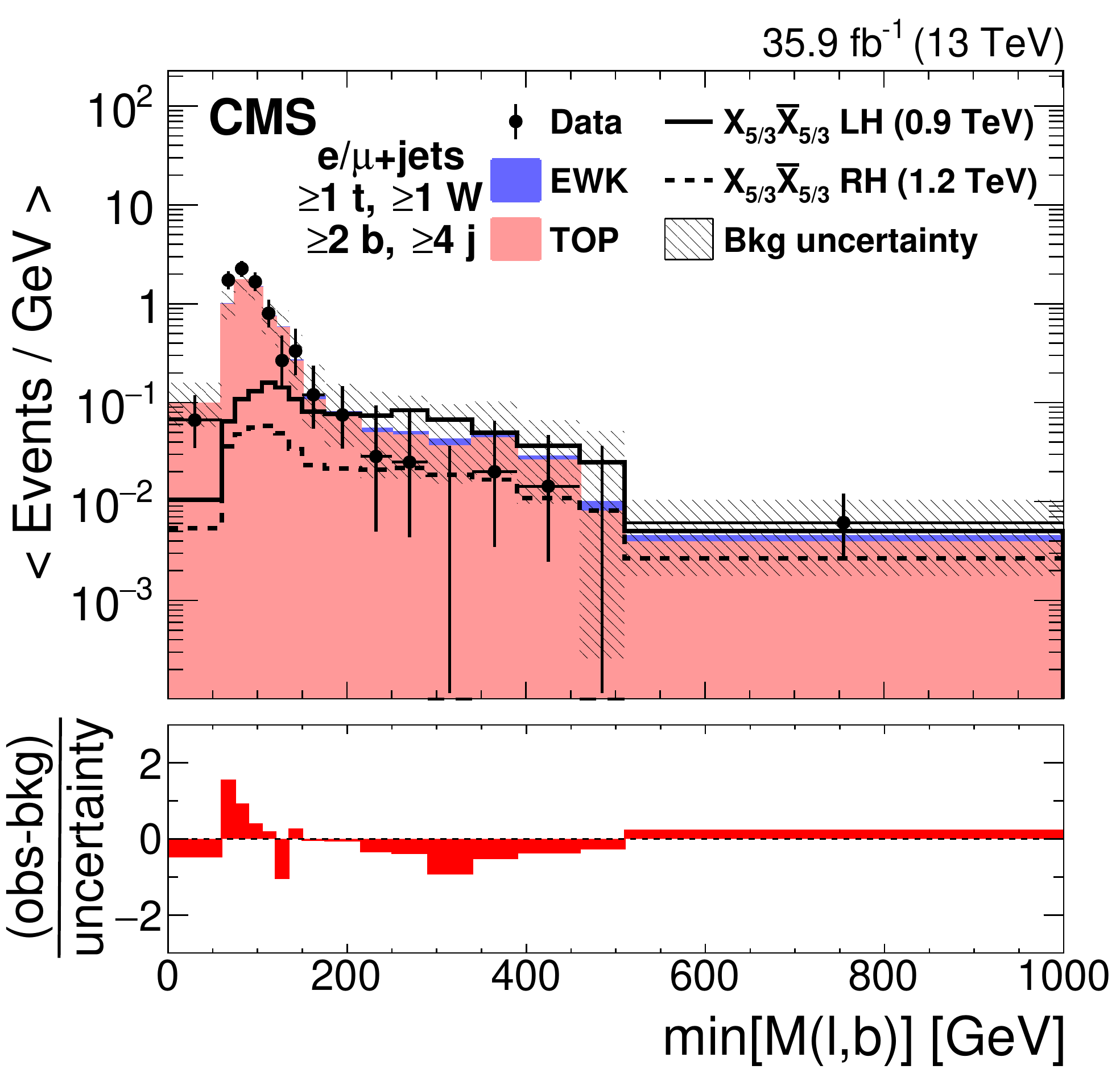}
\caption{Distributions of \minmlb in events with $\geq$1 {\cPqt}-tagged jets, 0 (upper) or $\geq$1 (lower) {\PW}-tagged jets,
and 1 (left) or $\geq$2 (right) {\cPqb}-tagged jets for the combined electron and muon samples in the signal region.
Example signal distributions are also shown.
The background distributions correspond to the background-only fit to data, while signal distributions are before the fit to data.
The last bin includes overflow events and its content is divided by the bin width.
The distributions in each category have variable-size bins,
chosen so that the statistical uncertainty in the total background in each bin is less than 30\%.
The lower panel in each plot shows the difference between the observed and the predicted numbers of events
in that bin divided by the total uncertainty.
The total uncertainty is calculated as the sum in quadrature of the statistical uncertainty
in the observed measurement and the statistical and systematic uncertainties in the background-only fit to data.}
\label{fig:Mlb1ptcats}
\end{figure}

In the single-lepton signal region, the LH (RH) signal efficiencies range between 4.1 (5.0) and 13.1 (14.7)\%.
Events in the signal region are separated into 16 categories based on the flavor of the lepton (\Pe,~\PGm), the number of {\cPqb}-tagged jets (1,~$\geq$2), the number of {\PW}-tagged jets (0,~$\geq$1), and the number of {\cPqt}-tagged jets (0,~$\geq$1).
Event yields for each analysis category are given in Table~\ref{tab:neventsfinal} after a background-only fit to data with the contribution from the electron and muon channels combined.
Figure~\ref{fig:Mlb0tcats} shows the distribution for \minmlb for events with zero {\cPqt}-tagged jets, while Fig.~\ref{fig:Mlb1ptcats} shows the \minmlb distribution for events with at least one {\cPqt}-tagged jet, both of which are shown after a background-only fit to data.
The distributions are separated for each analysis category, but again the electron and muon channels are combined.
No significant discrepancy is seen between the observed and predicted \minmlb distributions.

\begin{table}[htbp]
\centering
\topcaption{Expected (observed) numbers of background (data) events passing the final selection requirements,
in the signal region ($\drlj>1.0$) categories, after combining the single-electron and single-muon channels.
The numbers of events expected from two example signals are also shown.
The event yields and their uncertainties correspond to the background-only fit to data for the background, while for the signal they are based on the values before the fit to data.}
\begin{tabular}{l r@{\,$\pm$\,}l cr@{\,$\pm$\,}l cr@{\,$\pm$\,}l cr@{\,$\pm$\,}l}
Sample
& \multicolumn{2}{c}{0 {\cPqt}, 0 {\PW}, 1 {\cPqb}}
& \multicolumn{3}{c}{0 {\cPqt}, 0 {\PW}, $\geq$2 {\cPqb}}
& \multicolumn{3}{c}{0 {\cPqt}, $\geq$1 {\PW}, 1 {\cPqb}}
& \multicolumn{3}{c}{0 {\cPqt}, $\geq$1 {\PW}, $\geq$2 {\cPqb}} \\
\hline
LH \xft (0.9\TeV)    & 5.6 & 1.3         && 4.9 & 1.2         && 43.6 & 2.3       && 36.5 & 2.3   \rule{0pt}{2.5ex}  \\
RH \xft (1.2\TeV)    & 1.13 & 0.30       && 0.85 & 0.24       && 10.44 & 0.66     && 7.67 & 0.56    \\
[\cmsTabSkip]
TOP                  & 545 & 49          && 334 & 32          && 462 & 44         && 306 & 30       \\
EWK              	 & 366 & 27          && 54.0 & 4.7        && 108.5 & 9.3      && 19.7 & 2.7       \\
QCD         		 & 24.6 & 7.6        && 7.9 & 3.7         && 7.6 & 4.4        && 0.65 & $^{0.71}_{0.65}$       \\
[\cmsTabSkip]
Total bkg.    		 & 935 & 62          && 396 & 33          && 578 & 47         && 327 & 30       \\
Data           		 & \multicolumn{2}{c}{984} & \multicolumn{3}{c}{416}   & \multicolumn{3}{c}{577}  & \multicolumn{3}{c}{321}  \\
[\cmsTabSkip]
Sample
& \multicolumn{2}{c}{$\geq$1 {\cPqt}, 0 {\PW}, 1 {\cPqb}}
& \multicolumn{3}{c}{$\geq$1 {\cPqt}, 0 {\PW}, $\geq$2 {\cPqb}}
& \multicolumn{3}{c}{$\geq$1 {\cPqt}, $\geq$1 {\PW}, 1 {\cPqb}}
& \multicolumn{3}{c}{$\geq$1 {\cPqt}, $\geq$1 {\PW}, $\geq$2 {\cPqb}} \\
\hline
LH \xft (0.9\TeV)    & 17.6 & 1.6        && 15.5 & 1.5        && 39.7 & 2.3       && 34.5 & 2.2     \rule{0pt}{2.5ex}    \\
RH \xft (1.2\TeV)    & 4.16 & 0.52       && 3.40 & 0.49       && 13.82 & 0.84     && 11.83 & 0.82    \\
[\cmsTabSkip]
TOP               	 & 367 & 41          && 267 & 31          && 139 & 16         && 108 & 13        \\
EWK              	 & 108.7 & 9.0       && 19.3 & 1.8        && 22.6 & 3.6       && 2.69 & 0.31     \\
QCD            		 & 6.6 & 2.4         && 1.41 & 0.65       && 1.36 & 0.66      && 0.47 & 0.32     \\
[\cmsTabSkip]
Total bkg.     		 & 482 & 44          && 287 & 31          && 163 & 17         && 111 & 13        \\
Data        		 & \multicolumn{2}{c}{465} & \multicolumn{3}{c}{285}   & \multicolumn{3}{c}{135} & \multicolumn{3}{c}{123}   \\
\end{tabular}
\label{tab:neventsfinal}
\end{table}

\section{Systematic uncertainties}\label{sec:Systematics}

The uncertainties in the lepton reconstruction, identification, and isolation efficiencies are derived from the uncertainties in the data-to-simulation scale factors and range from 1 to 3\%.
These uncertainties are applied per lepton.
A 2.5\% uncertainty is assigned to the integrated luminosity measurement~\cite{CMS-PAS-LUM-17-001} used to scale the simulated signal and background processes.
The above uncertainties only affect the normalization of the simulated processes and not their shape.

Both final states have uncertainties in their simulation-based predictions from the uncertainties in the lepton triggering efficiency, the jet energy scale (JES), the jet energy resolution (JER), the pileup modeling, the cross section normalization, and the choice of PDFs.
For the same-sign dilepton final state, the uncertainty in the lepton triggering efficiency is 3\% while for the single-lepton final state it ranges between 2 and 5\%.
In both final states, this uncertainty is applied per event.
The JES and JER uncertainties are estimated by varying the relevant parameters up and down by one standard deviation (s.d.) and repeating the analysis selections.
The pileup uncertainty is assessed by varying the total inelastic cross section ($\sigma_{\text{inel.}}$) used in the pileup reweighting by $\pm$4.6\%~\cite{pileup2016}.
The uncertainty in the theoretical cross section from renormalization and factorization energy scales is estimated by independently varying the scales up and down by a factor of two and taking the maximum variation as the uncertainty.
The uncertainty associated with the PDFs used for the MC generation is evaluated from the set of NNPDF3.0 fitted replicas, following the standard procedure~\cite{NNPDF30}.

The single-lepton final state considers the shape variations in the signal distributions that come from varying the renormalization and factorization scales and the choice of PDF set.
For the same-sign dilepton final state, only their effect on the signal acceptance is considered, since a ``cut-and-count'' analysis is used in this case.
The normalization changes due to the variations in the signal acceptance are found to be negligible in the single-lepton final state.
The details of the systematic uncertainties are shown in Table~\ref{tab:MCUncert} for the same-sign dilepton final state and in Table~\ref{tab:sys-error} for the single-lepton final state.

\begin{table}[htbp]
\centering
\topcaption{Systematic uncertainties in percentage (\%) in the same-sign dilepton final state, associated with the simulated processes. The ``Normalization'' column refers to the uncertainties from the cross section normalization and the choice of PDF set.}
\begin{tabular}{lcccc}
Process       & JES & JER & Pileup & Normalization \\
\hline
$\ttbar\PW$   & 3 & 2 & 4    & 19 \\
$\ttbar\PZ$   & 3 & 2 & 4    & 12 \\
$\ttbar\PH$   & 3 & 2 & 4    & 30 \\
$\ttbar\ttbar$& 2 & 2 & 4    & 50 \\
\PW\cPZ       & 9 & 2 & 4    & 24 \\
\cPZ\cPZ      & 4 & 2 & 4    & 10 \\
$\PW\PW$      & 9 & 2 & 4    & 50 \\
$\PW\PW\PZ$   & 9 & 2 & 4    & 50 \\
$\PW\PZ\PZ$   & 9 & 2 & 4    & 50 \\
$\PZ\PZ\PZ$   & 9 & 2 & 4    & 50 \\
\xft          & 3 & 1 & 1    & \NA  \\
\end{tabular}
\label{tab:MCUncert}
\end{table}

In the single-lepton final state, uncertainties are also applied for the corrections on the {\cPqb} tagging, light quark mistagging, {\PW} tagging, and {\cPqt} tagging scale factors. The {\PW} tagging uncertainties have different components, which are treated as uncorrelated: corrections to the groomed mass scale and smearing, $\tau_{2}/\tau_{1}$ selection efficiency, and its \pt dependence.
For the top quark \pt reweighting, the difference between the weighted and unweighted distributions is added as a one-sided systematic uncertainty.

\begin{table}[htbp]
\centering
\topcaption{Summary of systematic uncertainties in the single-lepton final state. These uncertainties are included in both signal and all background processes, except for the top \pt systematic uncertainty, which is included only in \ttbar.
The range of uncertainty values in percentage (\%) corresponds to the effect on the yields before the fit to data and is given across the relevant background processes and channels for each systematic uncertainty.}
\begin{tabular}{l ccr@{--}lcc}
Source                         & \multicolumn{6}{c}{Uncertainty range} \\
\hline
Trigger efficiency             &&& 2&5   &&  \rule{0pt}{2.5ex} \\
Jet energy scale               &&& 0.5&52 && \\
Jet energy resolution          &&& 0&3 && \\
{\cPqb}/{\cPqc} tagging        &&& 0&5 && \\
{\cPqu\cPqd\cPqs\cPg} mistagging   &&& 0&4 && \\
{\PW} tagging: mass resolution     &&& 0&13 && \\
{\PW} tagging: mass scale          &&& 0&21 && \\
{\PW} tagging: $\tau_{2}/\tau_{1}$ &&& 0&2 && \\
{\PW} tagging: $\tau_{2}/\tau_{1}$ extrapolation &&& 0&2 && \\
{\cPqt} tagging                &&& 0&4 && \\
Top \pt                        &&& 0&19 && \\
Pileup                         &&& 0&4 && \\
PDF                            &&& 2&9 && \\
QCD renorm./fact. scale        &&& 12&36 && \\
\end{tabular}
\label{tab:sys-error}
\end{table}

Lastly, in the same-sign dilepton final state, there are uncertainties in the predictions of background processes whose estimates are made using control samples in data.
As stated above, a 30\% uncertainty is assigned to the predicted yield of background events from charge misidentification, and a 50\% uncertainty is assigned to the predicted yield of background events from processes with nonprompt leptons.

Systematic uncertainties that have the same source for the two different final states (\eg the uncertainty in the lepton identification) are treated as fully correlated between the two final states.

\section{Results}\label{sec:Results}

\begin{figure}[hbtp]
\centering
\includegraphics[width=0.49\textwidth]{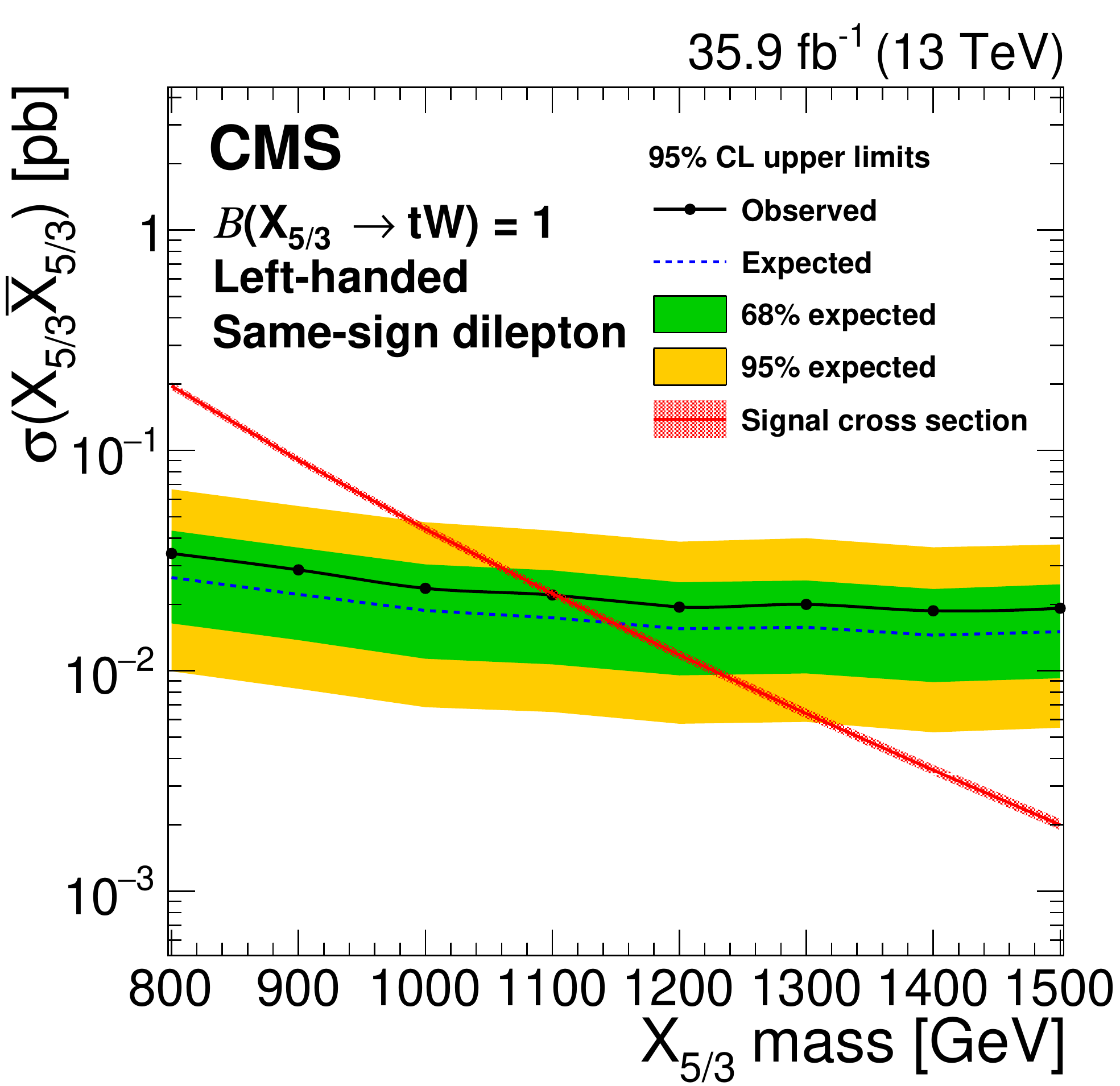}
\includegraphics[width=0.49\textwidth]{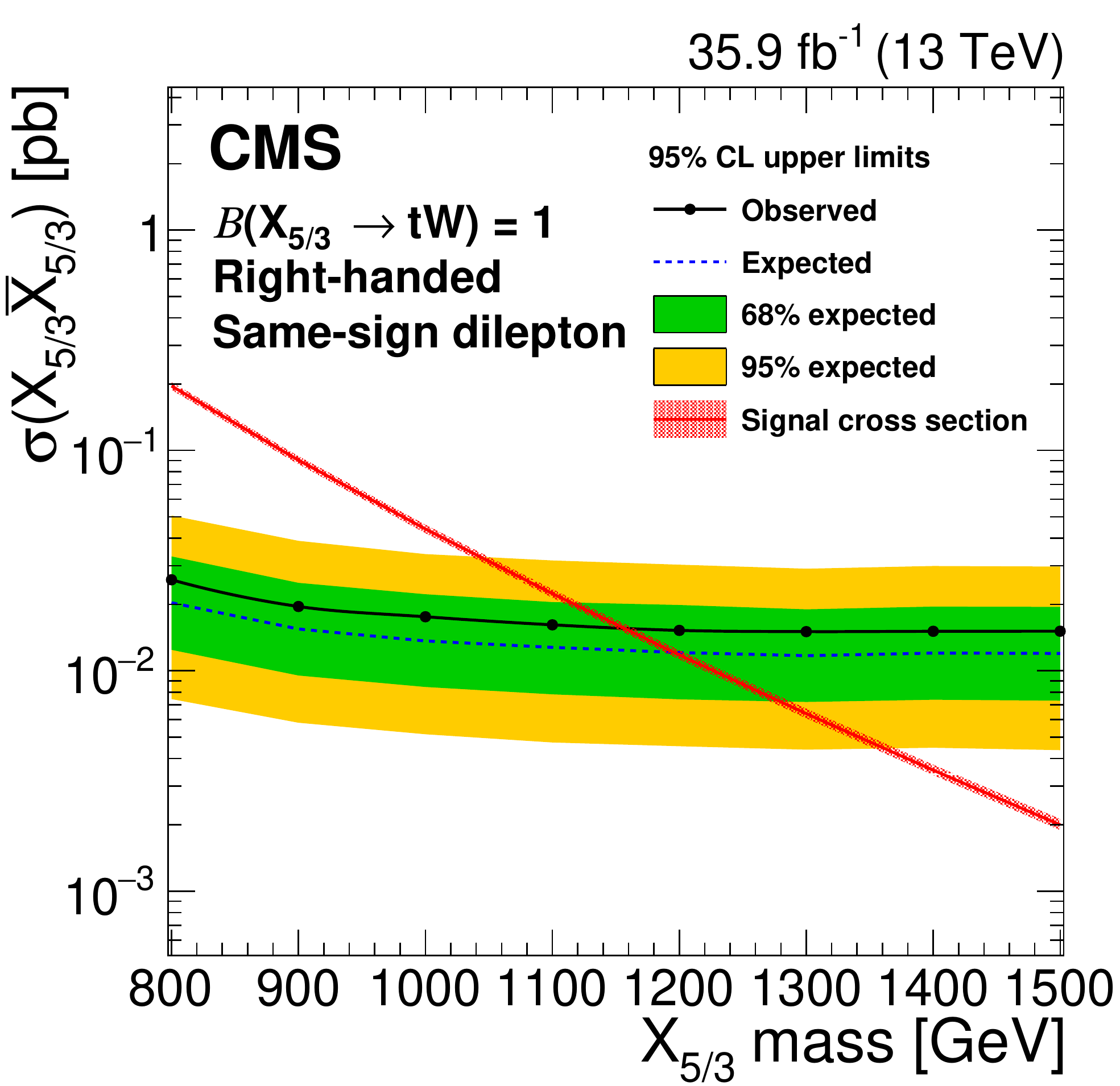}
\includegraphics[width=0.49\textwidth]{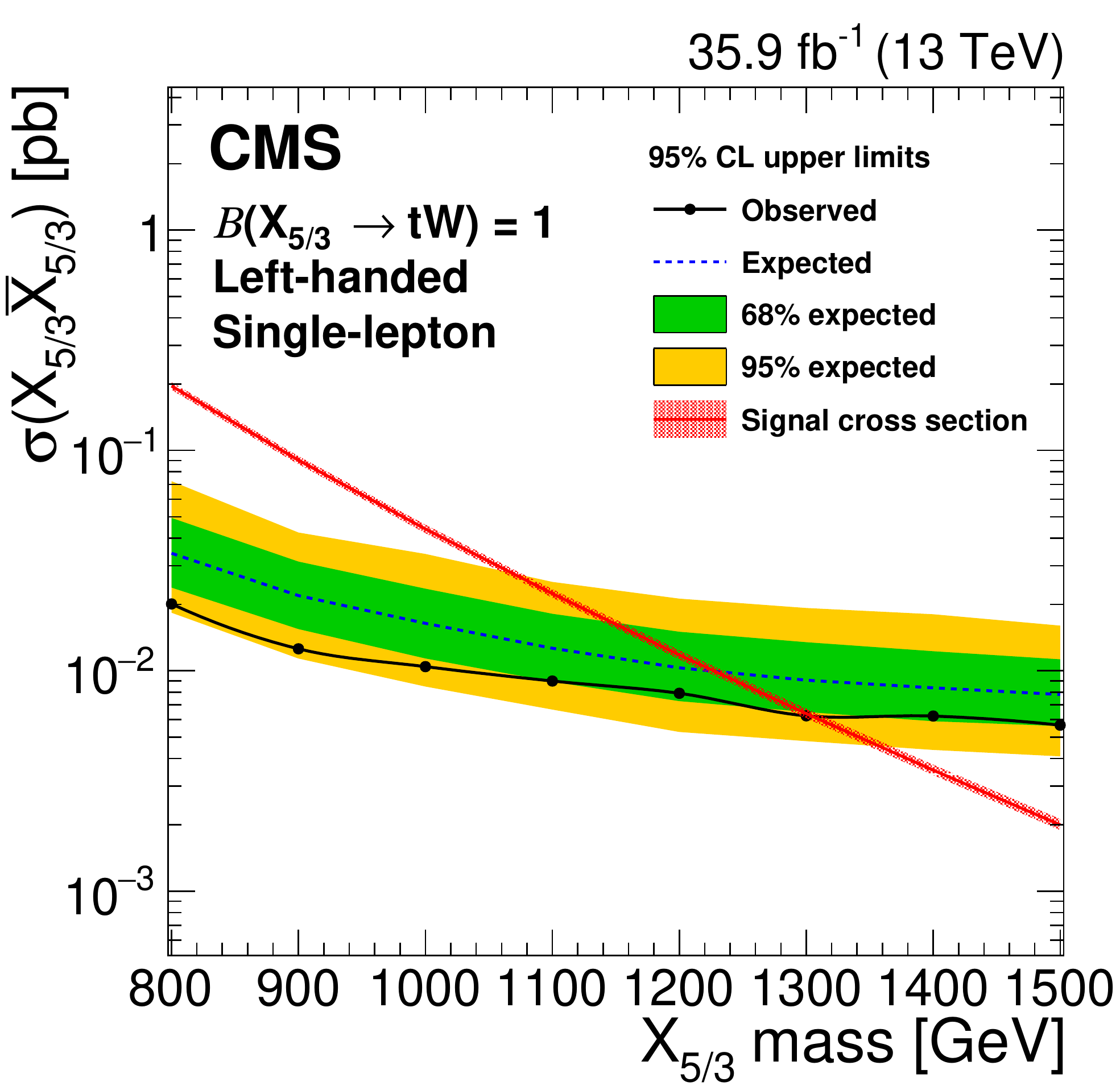}
\includegraphics[width=0.49\textwidth]{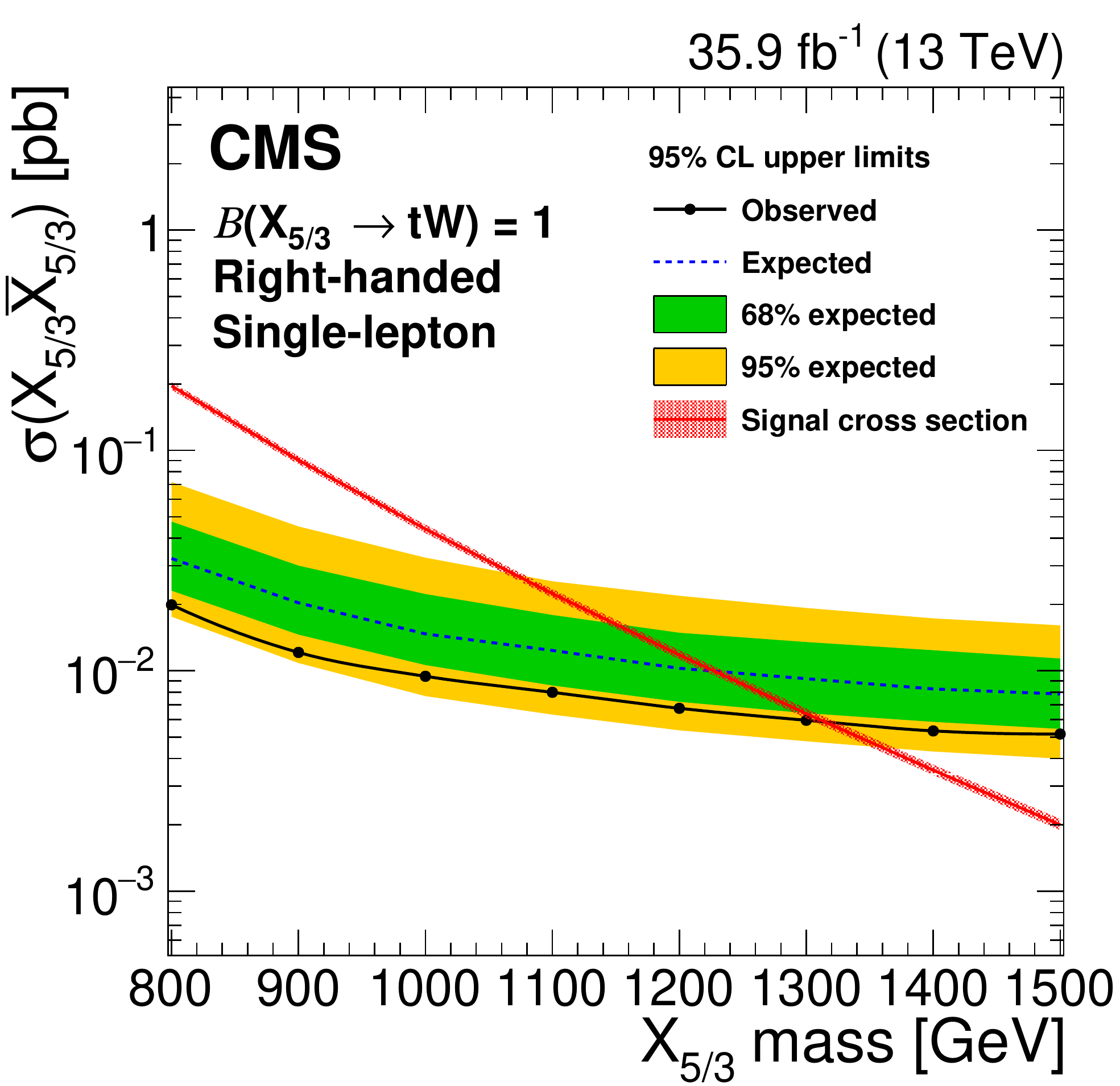}
\caption{Expected and observed limits at 95\% \CL for an LH (left) and RH (right) \xft after combining all categories for the same-sign dilepton (upper row) and the single-lepton (lower row) final states.
The theoretical uncertainty in the signal cross section is shown as a narrow band around the theoretical prediction.}
\label{fig:IndividualLimits}
\end{figure}

\begin{figure}[hbtp]
\centering
\includegraphics[width=0.49\textwidth]{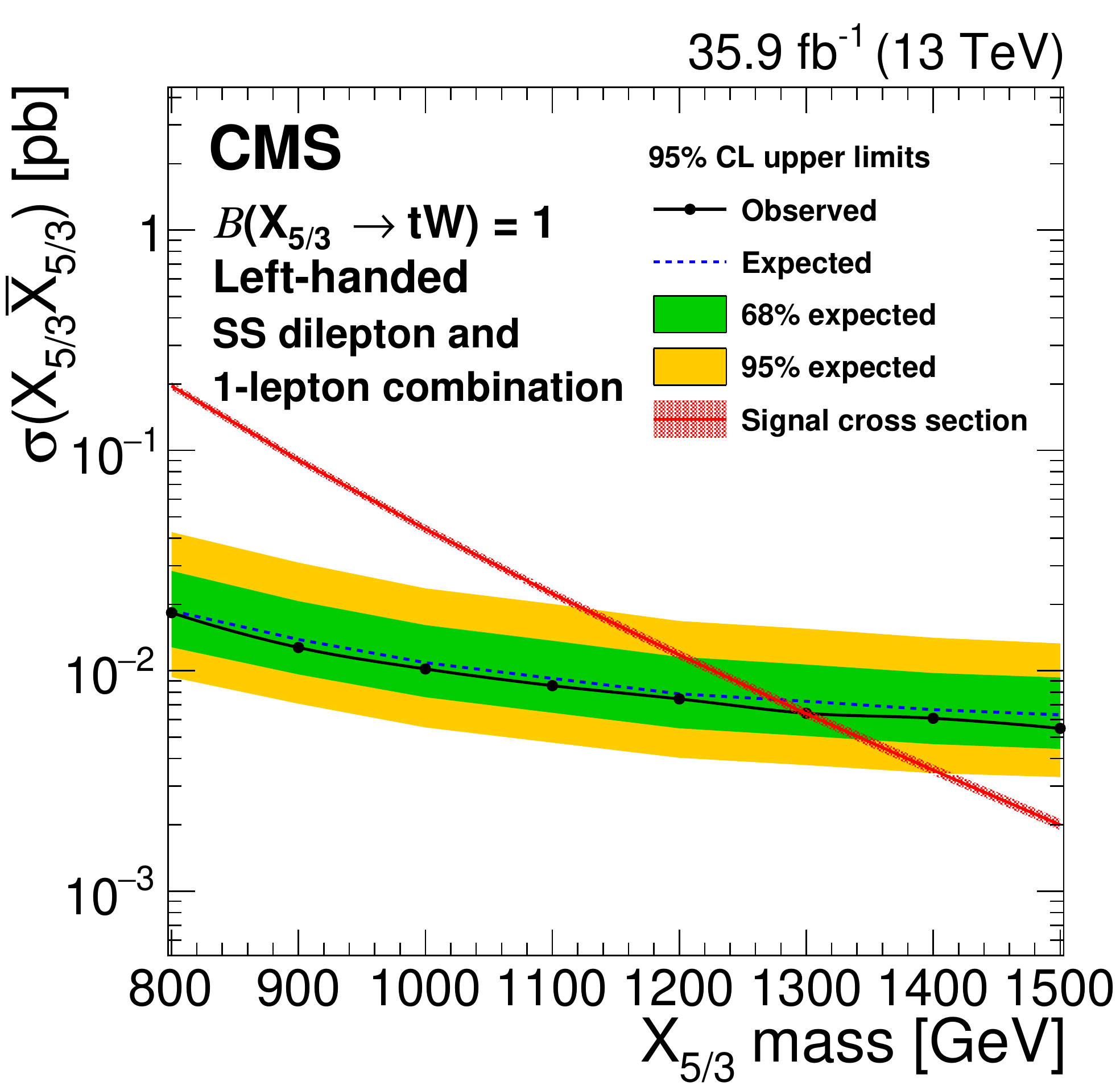}
\includegraphics[width=0.49\textwidth]{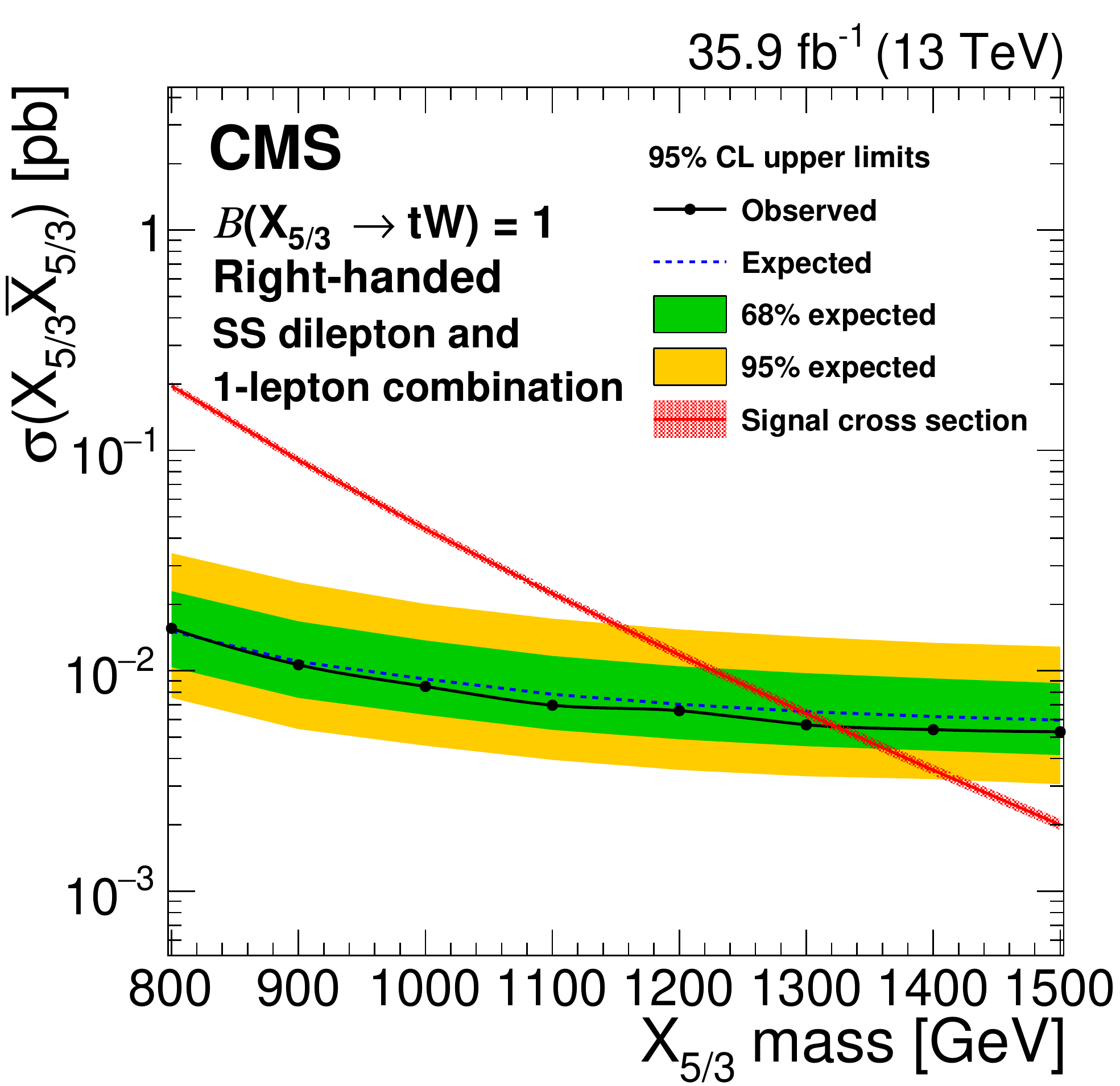}
\caption{Expected and observed limits at 95\% \CL for an LH (left) and RH (right) \xft after combining the same-sign dilepton and single-lepton final states.
The theoretical uncertainty in the signal cross section is shown as a narrow band around the theoretical prediction.}
\label{fig:CombinedLimits}
\end{figure}

No significant excess of events is observed above the SM prediction. Upper limits at 95\% \CL are set on the production cross sections $\Pp\Pp\to\xft \overline{\ensuremath{X}}_{5/3}$ for both couplings and for the different final states, as well as for their combination.
Bayesian statistics~\cite{CowanPDGStat,THETA} are used to calculate observed and expected limits with a flat prior taken for the signal cross section.
The same-sign dilepton final state limits are based on a counting experiment, while in the single-lepton final state, a binned likelihood fit on the distributions of \minmlb is performed simultaneously in the signal and control regions.
Systematic uncertainties are treated as nuisance parameters with normalization uncertainties having a log-normal prior and shape uncertainties a Gaussian prior.
The fit does not change any nuisance parameter by a significant amount compared to its pre-fit value.
After the full analysis selection described above, lower observed\,(expected) limits of 1.16\,(1.20) and 1.10\,(1.16)\TeV are placed on the mass of the \xft particle with RH and LH couplings to {\PW} bosons, respectively, using the same-sign dilepton final state.
In the single-lepton final state, observed (expected) limits of 1.32\,(1.23) and 1.30\,(1.23)\TeV are placed on the mass of the \xft particle with RH and LH couplings to {\PW} bosons, respectively.
Combining the two final states yields a lower observed (expected) limit on the \xft mass of 1.33 (1.30)\TeV for an \xft particle with RH couplings to {\PW} bosons and 1.30 (1.28)\TeV for an \xft particle with LH couplings to {\PW} bosons.
Figure~\ref{fig:IndividualLimits} shows the limits for the individual final states, while Fig.~\ref{fig:CombinedLimits} shows the limits obtained by combining the two final states.

\section{Summary}\label{sec:Summary}

A search has been performed for a heavy top quark partner with an exotic 5/3 charge (\xft) using proton-proton collision data collected by the CMS experiment in 2016 at a center-of-mass energy of 13\TeV and corresponding to 35.9\fbinv.
The \xft quark is assumed always to decay into a top quark and a {\PW} boson.
Two different final states, same-sign dilepton and single-lepton, are analyzed separately and then combined.
No significant excess over the expected standard model backgrounds is seen in data. Lower limits are set on the mass of the \xft particle. The observed (expected) limit is 1.33\,(1.30)\TeV for an \xft particle with right-handed couplings to {\PW} bosons and 1.30\,(1.28)\TeV for an \xft particle with left-handed couplings to {\PW} bosons in a combination of the same-sign dilepton and single-lepton final states.

\begin{acknowledgments}
We congratulate our colleagues in the CERN accelerator departments for the excellent performance of the LHC and thank the technical and administrative staffs at CERN and at other CMS institutes for their contributions to the success of the CMS effort. In addition, we gratefully acknowledge the computing centres and personnel of the Worldwide LHC Computing Grid for delivering so effectively the computing infrastructure essential to our analyses. Finally, we acknowledge the enduring support for the construction and operation of the LHC and the CMS detector provided by the following funding agencies: BMWFW and FWF (Austria); FNRS and FWO (Belgium); CNPq, CAPES, FAPERJ, and FAPESP (Brazil); MES (Bulgaria); CERN; CAS, MoST, and NSFC (China); COLCIENCIAS (Colombia); MSES and CSF (Croatia); RPF (Cyprus); SENESCYT (Ecuador); MoER, ERC IUT, and ERDF (Estonia); Academy of Finland, MEC, and HIP (Finland); CEA and CNRS/IN2P3 (France); BMBF, DFG, and HGF (Germany); GSRT (Greece); NKFIA (Hungary); DAE and DST (India); IPM (Iran); SFI (Ireland); INFN (Italy); MSIP and NRF (Republic of Korea); LAS (Lithuania); MOE and UM (Malaysia); BUAP, CINVESTAV, CONACYT, LNS, SEP, and UASLP-FAI (Mexico); MBIE (New Zealand); PAEC (Pakistan); MSHE and NSC (Poland); FCT (Portugal); JINR (Dubna); MON, RosAtom, RAS and RFBR (Russia); MESTD (Serbia); SEIDI, CPAN, PCTI and FEDER (Spain); Swiss Funding Agencies (Switzerland); MST (Taipei); ThEPCenter, IPST, STAR, and NSTDA (Thailand); TUBITAK and TAEK (Turkey); NASU and SFFR (Ukraine); STFC (United Kingdom); DOE and NSF (USA).

\hyphenation{Rachada-pisek} Individuals have received support from the Marie-Curie programme and the European Research Council and Horizon 2020 Grant, contract No. 675440 (European Union); the Leventis Foundation; the A. P. Sloan Foundation; the Alexander von Humboldt Foundation; the Belgian Federal Science Policy Office; the Fonds pour la Formation \`a la Recherche dans l'Industrie et dans l'Agriculture (FRIA-Belgium); the Agentschap voor Innovatie door Wetenschap en Technologie (IWT-Belgium); the F.R.S.-FNRS and FWO (Belgium) under the ``Excellence of Science - EOS" - be.h project n. 30820817; the Ministry of Education, Youth and Sports (MEYS) of the Czech Republic; the Lend\"ulet (``Momentum") Programme and the J\'anos Bolyai Research Scholarship of the Hungarian Academy of Sciences, the New National Excellence Program \'UNKP, the NKFIA research grants 123842, 123959, 124845, 124850 and 125105 (Hungary); the Council of Science and Industrial Research, India; the HOMING PLUS programme of the Foundation for Polish Science, cofinanced from European Union, Regional Development Fund, the Mobility Plus programme of the Ministry of Science and Higher Education, the National Science Center (Poland), contracts Harmonia 2014/14/M/ST2/00428, Opus 2014/13/B/ST2/02543, 2014/15/B/ST2/03998, and 2015/19/B/ST2/02861, Sonata-bis 2012/07/E/ST2/01406; the National Priorities Research Program by Qatar National Research Fund; the Programa Estatal de Fomento de la Investigaci{\'o}n Cient{\'i}fica y T{\'e}cnica de Excelencia Mar\'{\i}a de Maeztu, grant MDM-2015-0509 and the Programa Severo Ochoa del Principado de Asturias; the Thalis and Aristeia programmes cofinanced by EU-ESF and the Greek NSRF; the Rachadapisek Sompot Fund for Postdoctoral Fellowship, Chulalongkorn University and the Chulalongkorn Academic into Its 2nd Century Project Advancement Project (Thailand); the Welch Foundation, contract C-1845; and the Weston Havens Foundation (USA).
\end{acknowledgments}
\bibliography{auto_generated}

\providecommand{\href}[2]{#2}\begingroup\raggedright\begin{thebibliography}{10}%
\makeatletter
\providecommand{\hrefCMSnoop }[0]{\@secondoftwo}%
\makeatother
\providecommand{\doi}{\texttt{doi:}\begingroup \urlstyle{tt}\Url}

\bibitem{Contino:2008hi}
\hrefCMSnoop {}{R.~Contino and G.~Servant, ``Discovering the top partners at
  the {LHC} using same-sign dilepton final states'',} \textit{ JHEP} \textbf{
  06} (2008) 026,
  \href{http://dx.doi.org/10.1088/1126-6708/2008/06/026}{\doi{10.1088/1126-6708/2008/06/026}},
\href{http://www.arXiv.org/abs/0801.1679}{\texttt{arXiv:0801.1679}}.

\bibitem{TopPartnerHunterGuide}
\hrefCMSnoop {}{A.~De~Simone, O.~Matsedonskyi, R.~Rattazzi, and A.~Wulzer, ``A
  first top partner hunter's guide'',} \textit{ JHEP} \textbf{ 04} (2013) 004,
  \href{http://dx.doi.org/10.1007/JHEP04(2013)004}{\doi{10.1007/JHEP04(2013)004}},
\href{http://www.arXiv.org/abs/1211.5663}{\texttt{arXiv:1211.5663}}.

\bibitem{Pomarol}
\hrefCMSnoop {}{K.~Agashe, R.~Contino, and A.~Pomarol, ``The minimal composite
  {Higgs} model'',} \textit{ Nucl. Phys. B} \textbf{ 719} (2005) 165,
  \href{http://dx.doi.org/10.1016/j.nuclphysb.2005.04.035}{\doi{10.1016/j.nuclphysb.2005.04.035}},
\href{http://www.arXiv.org/abs/hep-ph/0412089}{\texttt{arXiv:hep-ph/0412089}}.

\bibitem{Kaplan}
\hrefCMSnoop {}{D.~B. Kaplan, ``Flavor at {SSC} energies: {A} new mechanism for
  dynamically generated fermion masses'',} \textit{ Nucl. Phys. B} \textbf{
  365} (1991) 259,
\href{http://dx.doi.org/10.1016/S0550-3213(05)80021-5}{\doi{10.1016/S0550-3213(05)80021-5}}.

\bibitem{Azatov:2011qy}
\hrefCMSnoop {}{A.~Azatov and J.~Galloway, ``Light custodians and {Higgs}
  physics in composite models'',} \textit{ Phys. Rev. D} \textbf{ 85} (2012)
  055013,
  \href{http://dx.doi.org/10.1103/PhysRevD.85.055013}{\doi{10.1103/PhysRevD.85.055013}},
\href{http://www.arXiv.org/abs/1110.5646}{\texttt{arXiv:1110.5646}}.

\bibitem{Cacciapaglia}
G.~Cacciapaglia\hrefCMSnoop {}{ {et~al.}, ``Heavy vector-like quark with charge
  5/3 at the {LHC}'',} \textit{ JHEP} \textbf{ 03} (2013) 004,
  \href{http://dx.doi.org/10.1007/JHEP03(2013)004}{\doi{10.1007/JHEP03(2013)004}},
\href{http://www.arXiv.org/abs/1211.4034}{\texttt{arXiv:1211.4034}}.

\bibitem{8TeVPaper}
\hrefCMSnoop {}{{CMS Collaboration}, ``Search for top-quark partners with
  charge 5/3 in the same-sign dilepton final state'',} \textit{ Phys. Rev.
  Lett.} \textbf{ 112} (2014) 171801,
  \href{http://dx.doi.org/10.1103/PhysRevLett.112.171801}{\doi{10.1103/PhysRevLett.112.171801}},
\href{http://www.arXiv.org/abs/1312.2391}{\texttt{arXiv:1312.2391}}.

\bibitem{B2G-15-006}
\hrefCMSnoop {}{{CMS Collaboration}, ``Search for top quark partners with
  charge 5/3 in proton-proton collisions at $\sqrt{s}=13$ {TeV}'',} \textit{
  JHEP} \textbf{ 08} (2017) 073,
  \href{http://dx.doi.org/10.1007/JHEP08(2017)073}{\doi{10.1007/JHEP08(2017)073}},
\href{http://www.arXiv.org/abs/1705.10967}{\texttt{arXiv:1705.10967}}.

\bibitem{Aad:2015gdg}
\hrefCMSnoop {}{{ATLAS Collaboration}, ``Analysis of events with $b$-jets and a
  pair of leptons of the same charge in $pp$ collisions at $\sqrt{s}=8$ {TeV}
  with the {ATLAS} detector'',} \textit{ JHEP} \textbf{ 10} (2015) 150,
  \href{http://dx.doi.org/10.1007/JHEP10(2015)150}{\doi{10.1007/JHEP10(2015)150}},
\href{http://www.arXiv.org/abs/1504.04605}{\texttt{arXiv:1504.04605}}.

\bibitem{Aad:2015mba}
\hrefCMSnoop {}{{ATLAS Collaboration}, ``Search for vector-like {$B$} quarks in
  events with one isolated lepton, missing transverse momentum and jets at
  $\sqrt{s}=8$ {TeV} with the {ATLAS} detector'',} \textit{ Phys. Rev. D}
  \textbf{ 91} (2015) 112011,
  \href{http://dx.doi.org/10.1103/PhysRevD.91.112011}{\doi{10.1103/PhysRevD.91.112011}},
\href{http://www.arXiv.org/abs/1503.05425}{\texttt{arXiv:1503.05425}}.

\bibitem{Aaboud:2017zfn}
\hrefCMSnoop {}{{ATLAS Collaboration}, ``Search for pair production of heavy
  vector-like quarks decaying to high-p$_{T}$ {W} bosons and b quarks in the
  lepton-plus-jets final state in pp collisions at $ \sqrt{s}=13$ {TeV} with
  the {ATLAS} detector'',} \textit{ JHEP} \textbf{ 10} (2017) 141,
  \href{http://dx.doi.org/10.1007/JHEP10(2017)141}{\doi{10.1007/JHEP10(2017)141}},
\href{http://www.arXiv.org/abs/1707.03347}{\texttt{arXiv:1707.03347}}.

\bibitem{Aaboud:2018uek}
\hrefCMSnoop {}{{ATLAS Collaboration}, ``Search for pair production of heavy
  vector-like quarks decaying into high-$\pt$ {$W$} bosons and top quarks in
  the lepton-plus-jets final state in $pp$ collisions at $\sqrt{s}=13$ {TeV}
  with the {ATLAS} detector'',} \textit{ JHEP} \textbf{ 08} (2018) 048,
  \href{http://dx.doi.org/10.1007/JHEP08(2018)048}{\doi{10.1007/JHEP08(2018)048}},
\href{http://www.arXiv.org/abs/1806.01762}{\texttt{arXiv:1806.01762}}.

\bibitem{Aaboud:2018xpj}
\hrefCMSnoop {}{{ATLAS Collaboration}, ``Search for new phenomena in events
  with same-charge leptons and $b$-jets in $pp$ collisions at $\sqrt{s}= 13$
  {TeV} with the {ATLAS} detector'',} \textit{ JHEP} \textbf{ 12} (2018) 039,
  \href{http://dx.doi.org/10.1007/JHEP12(2018)039}{\doi{10.1007/JHEP12(2018)039}},
\href{http://www.arXiv.org/abs/1807.11883}{\texttt{arXiv:1807.11883}}.

\bibitem{Aaboud:2018pii}
\hrefCMSnoop {}{{ATLAS Collaboration}, ``Combination of the searches for
  pair-produced vector-like partners of the third-generation quarks at
  $\sqrt{s}=$ 13 {TeV} with the {ATLAS} detector'',} \textit{ Phys. Rev. Lett.}
  \textbf{ 121} (2018) 211801,
  \href{http://dx.doi.org/10.1103/PhysRevLett.121.211801}{\doi{10.1103/PhysRevLett.121.211801}},
\href{http://www.arXiv.org/abs/1808.02343}{\texttt{arXiv:1808.02343}}.

\bibitem{Khachatryan:2016bia}
\hrefCMSnoop {}{{CMS Collaboration}, ``The {CMS} trigger system'',} \textit{
  JINST} \textbf{ 12} (2017) P01020,
  \href{http://dx.doi.org/10.1088/1748-0221/12/01/P01020}{\doi{10.1088/1748-0221/12/01/P01020}},
\href{http://www.arXiv.org/abs/1609.02366}{\texttt{arXiv:1609.02366}}.

\bibitem{Chatrchyan:2008zzk}
\hrefCMSnoop {}{{CMS Collaboration}, ``The {CMS} experiment at the {CERN}
  {LHC}'',} \textit{ JINST} \textbf{ 3} (2008) S08004,
\href{http://dx.doi.org/10.1088/1748-0221/3/08/S08004}{\doi{10.1088/1748-0221/3/08/S08004}}.

\bibitem{Alwall:2014hca}
J.~Alwall\hrefCMSnoop {}{ {et~al.}, ``The automated computation of tree-level
  and next-to-leading order differential cross sections, and their matching to
  parton shower simulations'',} \textit{ JHEP} \textbf{ 07} (2014) 079,
  \href{http://dx.doi.org/10.1007/JHEP07(2014)079}{\doi{10.1007/JHEP07(2014)079}},
  \href{http://www.arXiv.org/abs/1405.0301}{\texttt{arXiv:1405.0301}}.

\bibitem{MadSpin}
\hrefCMSnoop {}{P.~Artoisenet, R.~Frederix, O.~Mattelaer, and R.~Rietkerk,
  ``Automatic spin-entangled decays of heavy resonances in {Monte Carlo}
  simulations'',} \textit{ JHEP} \textbf{ 03} (2013) 015,
  \href{http://dx.doi.org/10.1007/JHEP03(2013)015}{\doi{10.1007/JHEP03(2013)015}},
\href{http://www.arXiv.org/abs/1212.3460}{\texttt{arXiv:1212.3460}}.

\bibitem{TPRIMEXSEC}
\hrefCMSnoop {}{M.~Czakon and A.~Mitov, ``Top++: a program for the calculation
  of the top-pair cross-section at hadron colliders'',} \textit{ Comput. Phys.
  Commun.} \textbf{ 185} (2014) 2930,
  \href{http://dx.doi.org/10.1016/j.cpc.2014.06.021}{\doi{10.1016/j.cpc.2014.06.021}},
\href{http://www.arXiv.org/abs/1112.5675}{\texttt{arXiv:1112.5675}}.

\bibitem{MITOV1}
\hrefCMSnoop {}{M.~Czakon, P.~Fiedler, and A.~Mitov, ``The total top quark pair
  production cross-section at hadron colliders through
  $\mathcal{O}({\ensuremath{\alpha}}_{S}^{4})$'',} \textit{ Phys. Rev. Lett.}
  \textbf{ 110} (2013) 252004,
  \href{http://dx.doi.org/10.1103/PhysRevLett.110.252004}{\doi{10.1103/PhysRevLett.110.252004}},
\href{http://www.arXiv.org/abs/1303.6254}{\texttt{arXiv:1303.6254}}.

\bibitem{MITOV3}
\hrefCMSnoop {}{M.~Czakon and A.~Mitov, ``{NNLO} corrections to top-pair
  production at hadron colliders: the all-fermionic scattering channels'',}
  \textit{ JHEP} \textbf{ 12} (2012) 054,
  \href{http://dx.doi.org/10.1007/JHEP12(2012)054}{\doi{10.1007/JHEP12(2012)054}},
\href{http://www.arXiv.org/abs/1207.0236}{\texttt{arXiv:1207.0236}}.

\bibitem{MITOV2}
\hrefCMSnoop {}{M.~Czakon and A.~Mitov, ``{NNLO} corrections to top pair
  production at hadron colliders: the quark-gluon reaction'',} \textit{ JHEP}
  \textbf{ 01} (2013) 080,
  \href{http://dx.doi.org/10.1007/JHEP01(2013)080}{\doi{10.1007/JHEP01(2013)080}},
\href{http://www.arXiv.org/abs/1210.6832}{\texttt{arXiv:1210.6832}}.

\bibitem{BARNREUTHER}
\hrefCMSnoop {}{P.~B{\"a}rnreuther, M.~Czakon, and A.~Mitov, ``Percent level
  precision physics at the {Tevatron}: first genuine {NNLO QCD} corrections to
  $q \bar{q} \to t \bar{t} + {X}$'',} \textit{ Phys. Rev. Lett.} \textbf{ 109}
  (2012) 132001,
  \href{http://dx.doi.org/10.1103/PhysRevLett.109.132001}{\doi{10.1103/PhysRevLett.109.132001}},
\href{http://www.arXiv.org/abs/1204.5201}{\texttt{arXiv:1204.5201}}.

\bibitem{NNLL}
M.~Cacciari\hrefCMSnoop {}{ {et~al.}, ``Top-pair production at hadron colliders
  with next-to-next-to-leading logarithmic soft-gluon resummation'',} \textit{
  Phys. Lett. B} \textbf{ 710} (2012) 612,
  \href{http://dx.doi.org/10.1016/j.physletb.2012.03.013}{\doi{10.1016/j.physletb.2012.03.013}},
\href{http://www.arXiv.org/abs/1111.5869}{\texttt{arXiv:1111.5869}}.

\bibitem{Nason:2004rx}
\hrefCMSnoop {}{P.~Nason, ``A new method for combining {NLO} {QCD} with shower
  {Monte Carlo} algorithms'',} \textit{ JHEP} \textbf{ 11} (2004) 040,
  \href{http://dx.doi.org/10.1088/1126-6708/2004/11/040}{\doi{10.1088/1126-6708/2004/11/040}},
\href{http://www.arXiv.org/abs/hep-ph/0409146}{\texttt{arXiv:hep-ph/0409146}}.

\bibitem{Frixione:2007vw}
\hrefCMSnoop {}{S.~Frixione, P.~Nason, and C.~Oleari, ``Matching {NLO} {QCD}
  computations with parton shower simulations: the {POWHEG} method'',} \textit{
  JHEP} \textbf{ 11} (2007) 070,
  \href{http://dx.doi.org/10.1088/1126-6708/2007/11/070}{\doi{10.1088/1126-6708/2007/11/070}},
\href{http://www.arXiv.org/abs/0709.2092}{\texttt{arXiv:0709.2092}}.

\bibitem{Alioli:2010xd}
\hrefCMSnoop {}{S.~Alioli, P.~Nason, C.~Oleari, and E.~Re, ``A general
  framework for implementing {NLO} calculations in shower {Monte Carlo}
  programs: the {POWHEG BOX}'',} \textit{ JHEP} \textbf{ 06} (2010) 043,
  \href{http://dx.doi.org/10.1007/JHEP06(2010)043}{\doi{10.1007/JHEP06(2010)043}},
\href{http://www.arXiv.org/abs/1002.2581}{\texttt{arXiv:1002.2581}}.

\bibitem{Frixione:2007nw}
\hrefCMSnoop {}{S.~Frixione, P.~Nason, and G.~Ridolfi, ``A positive-weight
  next-to-leading-order {Monte Carlo} for heavy flavour hadroproduction'',}
  \textit{ JHEP} \textbf{ 09} (2007) 126,
  \href{http://dx.doi.org/10.1088/1126-6708/2007/09/126}{\doi{10.1088/1126-6708/2007/09/126}},
\href{http://www.arXiv.org/abs/0707.3088}{\texttt{arXiv:0707.3088}}.

\bibitem{Alwall:2007fs}
J.~Alwall\hrefCMSnoop {}{ {et~al.}, ``Comparative study of various algorithms
  for the merging of parton showers and matrix elements in hadronic
  collisions'',} \textit{ Eur. Phys. J. C} \textbf{ 53} (2008) 473,
  \href{http://dx.doi.org/10.1140/epjc/s10052-007-0490-5}{\doi{10.1140/epjc/s10052-007-0490-5}},
\href{http://www.arXiv.org/abs/0706.2569}{\texttt{arXiv:0706.2569}}.

\bibitem{Frederix:2012ps}
\hrefCMSnoop {}{R.~Frederix and S.~Frixione, ``Merging meets matching in
  {MC@NLO}'',} \textit{ JHEP} \textbf{ 12} (2012) 061,
  \href{http://dx.doi.org/10.1007/JHEP12(2012)061}{\doi{10.1007/JHEP12(2012)061}},
\href{http://www.arXiv.org/abs/1209.6215}{\texttt{arXiv:1209.6215}}.

\bibitem{PYTHIA82}
T.~Sj{\"o}strand\hrefCMSnoop {}{ {et~al.}, ``An introduction to {PYTHIA
  8.2}'',} \textit{ Comput. Phys. Commun.} \textbf{ 191} (2015) 159,
  \href{http://dx.doi.org/10.1016/j.cpc.2015.01.024}{\doi{10.1016/j.cpc.2015.01.024}},
\href{http://www.arXiv.org/abs/1410.3012}{\texttt{arXiv:1410.3012}}.

\bibitem{NNPDF30}
\hrefCMSnoop {}{{NNPDF} Collaboration, ``Parton distributions for the {LHC Run
  II}'',} \textit{ JHEP} \textbf{ 04} (2015) 040,
  \href{http://dx.doi.org/10.1007/JHEP04(2015)040}{\doi{10.1007/JHEP04(2015)040}},
\href{http://www.arXiv.org/abs/1410.8849}{\texttt{arXiv:1410.8849}}.

\bibitem{CUETP8M1}
\hrefCMSnoop {}{{CMS Collaboration}, ``Event generator tunes obtained from
  underlying event and multiparton scattering measurements'',} \textit{ Eur.
  Phys. J. C} \textbf{ 76} (2016) 155,
  \href{http://dx.doi.org/10.1140/epjc/s10052-016-3988-x}{\doi{10.1140/epjc/s10052-016-3988-x}},
\href{http://www.arXiv.org/abs/1512.00815}{\texttt{arXiv:1512.00815}}.

\bibitem{Skands:2014pea}
\hrefCMSnoop {}{P.~Skands, S.~Carrazza, and J.~Rojo, ``Tuning {PYTHIA 8.1}: the
  {Monash} 2013 tune'',} \textit{ Eur. Phys. J. C} \textbf{ 74} (2014) 3024,
  \href{http://dx.doi.org/10.1140/epjc/s10052-014-3024-y}{\doi{10.1140/epjc/s10052-014-3024-y}},
\href{http://www.arXiv.org/abs/1404.5630}{\texttt{arXiv:1404.5630}}.

\bibitem{CMS-PAS-TOP-16-021}
\href {https://cds.cern.ch/record/2235192}{{CMS Collaboration},
  ``Investigations of the impact of the parton shower tuning in {Pythia 8} in
  the modelling of $\mathrm{t\overline{t}}$ at $\sqrt{s}=8$ and 13 {TeV}'',}
  CMS Physics Analysis Summary CMS-PAS-TOP-16-021, 2016.

\bibitem{GEANT}
\hrefCMSnoop {}{{GEANT4} Collaboration, ``{\GEANTfour}---a simulation
  toolkit'',} \textit{ Nucl. Instrum. Meth. A} \textbf{ 506} (2003) 250,
\href{http://dx.doi.org/10.1016/S0168-9002(03)01368-8}{\doi{10.1016/S0168-9002(03)01368-8}}.

\bibitem{TopPt}
\hrefCMSnoop {}{{CMS Collaboration}, ``Measurement of the differential cross
  section for top quark pair production in pp collisions at $\sqrt{s}=8$
  {TeV}'',} \textit{ Eur. Phys. J. C} \textbf{ 75} (2015) 542,
  \href{http://dx.doi.org/10.1140/epjc/s10052-015-3709-x}{\doi{10.1140/epjc/s10052-015-3709-x}},
\href{http://www.arXiv.org/abs/1505.04480}{\texttt{arXiv:1505.04480}}.

\bibitem{pf}
\hrefCMSnoop {}{{CMS Collaboration}, ``Particle-flow reconstruction and global
  event description with the {CMS} detector'',} \textit{ JINST} \textbf{ 12}
  (2017) P10003,
  \href{http://dx.doi.org/10.1088/1748-0221/12/10/P10003}{\doi{10.1088/1748-0221/12/10/P10003}},
\href{http://www.arXiv.org/abs/1706.04965}{\texttt{arXiv:1706.04965}}.

\bibitem{Cacciari:2008gp}
\hrefCMSnoop {}{M.~Cacciari, G.~P. Salam, and G.~Soyez, ``The anti-$\kt$ jet
  clustering algorithm'',} \textit{ JHEP} \textbf{ 04} (2008) 063,
  \href{http://dx.doi.org/10.1088/1126-6708/2008/04/063}{\doi{10.1088/1126-6708/2008/04/063}},
  \href{http://www.arXiv.org/abs/0802.1189}{\texttt{arXiv:0802.1189}}.

\bibitem{FastJet3}
\hrefCMSnoop {}{M.~Cacciari, G.~P. Salam, and G.~Soyez, ``{FastJet} user
  manual'',} \textit{ Eur. Phys. J. C} \textbf{ 72} (2012) 1896,
  \href{http://dx.doi.org/10.1140/epjc/s10052-012-1896-2}{\doi{10.1140/epjc/s10052-012-1896-2}},
\href{http://www.arXiv.org/abs/1111.6097}{\texttt{arXiv:1111.6097}}.

\bibitem{8TeV-EGamma}
\hrefCMSnoop {}{{CMS Collaboration}, ``Performance of electron reconstruction
  and selection with the {CMS} detector in proton-proton collisions at
  $\sqrt{s}=8$ {TeV}'',} \textit{ JINST} \textbf{ 10} (2015) P06005,
  \href{http://dx.doi.org/10.1088/1748-0221/10/06/P06005}{\doi{10.1088/1748-0221/10/06/P06005}},
\href{http://www.arXiv.org/abs/1502.02701}{\texttt{arXiv:1502.02701}}.

\bibitem{Khachatryan:1277738}
\hrefCMSnoop {}{{CMS Collaboration}, ``{CMS} tracking performance results from
  early {LHC} operation'',} \textit{ Eur. Phys. J. C} \textbf{ 70} (2010) 1165,
  \href{http://dx.doi.org/10.1140/epjc/s10052-010-1491-3}{\doi{10.1140/epjc/s10052-010-1491-3}},
\href{http://www.arXiv.org/abs/1007.1988}{\texttt{arXiv:1007.1988}}.

\bibitem{gsf}
\hrefCMSnoop {}{W.~Adam, R.~Fr{\"u}hwirth, A.~Strandlie, and T.~Todorov,
  ``Reconstruction of electrons with the {Gaussian-sum} filter in the {CMS}
  tracker at the {LHC}'',} \textit{ J. Phys. G} \textbf{ 31} (2005) N9,
  \href{http://dx.doi.org/10.1088/0954-3899/31/9/N01}{\doi{10.1088/0954-3899/31/9/N01}},
\href{http://www.arXiv.org/abs/physics/0306087}{\texttt{arXiv:physics/0306087}}.

\bibitem{FastJet1}
\hrefCMSnoop {}{M.~Cacciari and G.~P. Salam, ``Pileup subtraction using jet
  areas'',} \textit{ Phys. Lett. B} \textbf{ 659} (2008) 119,
  \href{http://dx.doi.org/10.1016/j.physletb.2007.09.077}{\doi{10.1016/j.physletb.2007.09.077}},
\href{http://www.arXiv.org/abs/0707.1378}{\texttt{arXiv:0707.1378}}.

\bibitem{CMS:2011aa}
\hrefCMSnoop {}{{CMS Collaboration}, ``Measurement of the inclusive {W} and {Z}
  production cross sections in pp collisions at $\sqrt{s}=7$ {TeV}'',} \textit{
  JHEP} \textbf{ 10} (2011) 132,
  \href{http://dx.doi.org/10.1007/JHEP10(2011)132}{\doi{10.1007/JHEP10(2011)132}},
\href{http://www.arXiv.org/abs/1107.4789}{\texttt{arXiv:1107.4789}}.

\bibitem{FastJet2}
\hrefCMSnoop {}{M.~Cacciari, G.~P. Salam, and G.~Soyez, ``The catchment area of
  jets'',} \textit{ JHEP} \textbf{ 04} (2008) 005,
  \href{http://dx.doi.org/10.1088/1126-6708/2008/04/005}{\doi{10.1088/1126-6708/2008/04/005}},
\href{http://www.arXiv.org/abs/0802.1188}{\texttt{arXiv:0802.1188}}.

\bibitem{CMS-PAS-JME-16-003}
\href {https://cds.cern.ch/record/2256875}{{CMS Collaboration}, ``Jet
  algorithms performance in 13 {TeV} data'',} CMS Physics Analysis Summary
  CMS-PAS-JME-16-003, 2017.

\bibitem{JECold}
\hrefCMSnoop {}{{CMS Collaboration}, ``Determination of jet energy calibration
  and transverse momentum resolution in {CMS}'',} \textit{ JINST} \textbf{ 6}
  (2011) P11002,
  \href{http://dx.doi.org/10.1088/1748-0221/6/11/P11002}{\doi{10.1088/1748-0221/6/11/P11002}},
\href{http://www.arXiv.org/abs/1107.4277}{\texttt{arXiv:1107.4277}}.

\bibitem{JEC}
\hrefCMSnoop {}{{CMS Collaboration}, ``Jet energy scale and resolution in the
  {CMS} experiment in {\Pp\Pp} collisions at 8 {TeV}'',} \textit{ JINST}
  \textbf{ 12} (2017) P02014,
  \href{http://dx.doi.org/10.1088/1748-0221/12/02/P02014}{\doi{10.1088/1748-0221/12/02/P02014}},
\href{http://www.arXiv.org/abs/1607.03663}{\texttt{arXiv:1607.03663}}.

\bibitem{btagRun2}
\hrefCMSnoop {}{{CMS Collaboration}, ``Identification of heavy-flavour jets
  with the {CMS} detector in pp collisions at 13 {TeV}'',} \textit{ JINST}
  \textbf{ 13} (2018) P05011,
  \href{http://dx.doi.org/10.1088/1748-0221/13/05/P05011}{\doi{10.1088/1748-0221/13/05/P05011}},
\href{http://www.arXiv.org/abs/1712.07158}{\texttt{arXiv:1712.07158}}.

\bibitem{softdrop}
\hrefCMSnoop {}{A.~J. Larkoski, S.~Marzani, G.~Soyez, and J.~Thaler, ``Soft
  drop'',} \textit{ JHEP} \textbf{ 05} (2014) 146,
  \href{http://dx.doi.org/10.1007/JHEP05(2014)146}{\doi{10.1007/JHEP05(2014)146}},
  \href{http://www.arXiv.org/abs/1402.2657}{\texttt{arXiv:1402.2657}}.

\bibitem{NSUBJETS}
\hrefCMSnoop {}{J.~Thaler and K.~Van~Tilburg, ``Maximizing boosted top
  identification by minimizing {$N$}-subjettiness'',} \textit{ JHEP} \textbf{
  02} (2012) 093,
  \href{http://dx.doi.org/10.1007/JHEP02(2012)093}{\doi{10.1007/JHEP02(2012)093}},
\href{http://www.arXiv.org/abs/1108.2701}{\texttt{arXiv:1108.2701}}.

\bibitem{pruning}
\hrefCMSnoop {}{S.~D. Ellis, C.~K. Vermilion, and J.~R. Walsh, ``Techniques for
  improved heavy particle searches with jet substructure'',} \textit{ Phys.
  Rev. D} \textbf{ 80} (2009) 051501,
  \href{http://dx.doi.org/10.1103/PhysRevD.80.051501}{\doi{10.1103/PhysRevD.80.051501}},
\href{http://www.arXiv.org/abs/0903.5081}{\texttt{arXiv:0903.5081}}.

\bibitem{susy2015}
\hrefCMSnoop {}{{CMS Collaboration}, ``Search for new physics in same-sign
  dilepton events in proton-proton collisions at $\sqrt{s}=13$ {TeV}'',}
  \textit{ Eur. Phys. J. C} \textbf{ 76} (2016) 439,
  \href{http://dx.doi.org/10.1140/epjc/s10052-016-4261-z}{\doi{10.1140/epjc/s10052-016-4261-z}},
\href{http://www.arXiv.org/abs/1605.03171}{\texttt{arXiv:1605.03171}}.

\bibitem{susy2011}
\hrefCMSnoop {}{{CMS Collaboration}, ``Search for new physics with same-sign
  isolated dilepton events with jets and missing transverse energy at the
  {LHC}'',} \textit{ JHEP} \textbf{ 06} (2011) 077,
  \href{http://dx.doi.org/10.1007/JHEP06(2011)077}{\doi{10.1007/JHEP06(2011)077}},
\href{http://www.arXiv.org/abs/1104.3168}{\texttt{arXiv:1104.3168}}.

\bibitem{CMS-PAS-LUM-17-001}
\href {https://cds.cern.ch/record/2257069}{{CMS Collaboration}, ``{CMS}
  luminosity measurement for the 2016 data-taking period'',} CMS Physics
  Analysis Summary CMS-PAS-LUM-17-001, 2017.

\bibitem{pileup2016}
\hrefCMSnoop {}{{CMS Collaboration}, ``Measurement of the inelastic
  proton-proton cross section at $\sqrt{s}=$ 13 {TeV}'',} \textit{ JHEP}
  \textbf{ 07} (2018) 161,
  \href{http://dx.doi.org/10.1007/JHEP07(2018)161}{\doi{10.1007/JHEP07(2018)161}},
\href{http://www.arXiv.org/abs/1802.02613}{\texttt{arXiv:1802.02613}}.

\bibitem{CowanPDGStat}
\hrefCMSnoop {}{{Particle Data Group}, M.~Tanabashi {et~al.}, ``Review of
  particle physics'',} \textit{ Phys. Rev. D} \textbf{ 98} (2018) 030001,
  \href{http://dx.doi.org/10.1103/PhysRevD.98.030001}{\doi{10.1103/PhysRevD.98.030001}}.
See {Ch. 39, ``Statistics'', G. Cowan}.

\bibitem{THETA}
\href {http://www-ekp.physik.uni-karlsruhe.de/\~ott/theta/theta-auto}{J.~Ott,
  ``\textsc{Theta}---{A} framework for template-based modeling and
  inference'',} 2010.
\newblock \url {http://www-ekp.physik.uni-karlsruhe.de/\~ott/theta/theta-auto}.

\end{thebibliography}\endgroup

\cleardoublepage \appendix\section{The CMS Collaboration \label{app:collab}}\begin{sloppypar}\hyphenpenalty=5000\widowpenalty=500\clubpenalty=5000\vskip\cmsinstskip
\textbf{Yerevan Physics Institute, Yerevan, Armenia}\\*[0pt]
A.M.~Sirunyan, A.~Tumasyan
\vskip\cmsinstskip
\textbf{Institut f\"{u}r Hochenergiephysik, Wien, Austria}\\*[0pt]
W.~Adam, F.~Ambrogi, E.~Asilar, T.~Bergauer, J.~Brandstetter, M.~Dragicevic, J.~Er\"{o}, A.~Escalante~Del~Valle, M.~Flechl, R.~Fr\"{u}hwirth\cmsAuthorMark{1}, V.M.~Ghete, J.~Hrubec, M.~Jeitler\cmsAuthorMark{1}, N.~Krammer, I.~Kr\"{a}tschmer, D.~Liko, T.~Madlener, I.~Mikulec, N.~Rad, H.~Rohringer, J.~Schieck\cmsAuthorMark{1}, R.~Sch\"{o}fbeck, M.~Spanring, D.~Spitzbart, A.~Taurok, W.~Waltenberger, J.~Wittmann, C.-E.~Wulz\cmsAuthorMark{1}, M.~Zarucki
% ...
\vskip\cmsinstskip
% ...
1:  Also at Vienna University of Technology, Vienna, Austria\\
% ...
\end{sloppypar}
\end{document}